\documentclass{aa}  

\usepackage{newtxtext}
\usepackage{color}
\usepackage{amsmath}
\usepackage{hyperref}
\usepackage{txfonts}
\usepackage{soul}

\begin{document}

   \title{On the inclinations of the Jupiter Trojans}

   \subtitle{}

   \author{Simona Pirani\inst{1}
          \and
          Anders Johansen\inst{1}
          \and
          Alexander J. Mustill\inst{1}  
          }

   \institute{Lund Observatory, Department of Astronomy and Theoretical Physics, Lund University, Box 43, 22100 Lund, Sweden.\\
              \email{simona@astro.lu.se}            
               }

   \date{}

 
  \abstract
  {Jupiter Trojans are a resonant asteroidal population characterised by photometric colours compatible with Trans-Neptunian objects, high inclinations and an asymmetric distribution of the number of asteroids between the two swarms. Different models have been proposed to explain the high inclination of the Trojans and to interpret their relation with the Trans-Neptunian objects, but none of these models can also satisfactorily explain the asymmetry ratio between the number of asteroids in the two swarms. Recently it has been found that the asymmetry ratio can arise if Jupiter has migrated inwards through the protoplanetary disc by at least a few astronomical units during its growth. The higher population of the leading swarm and the dark photometric colours of the Trojans are natural outcomes of this new model, but simulations with massless unperturbed disc particles led to a flat distribution of the Trojan inclinations and a final total mass of the Trojans that was 3-4 orders of magnitude larger than the current one. 
In our work, we investigate the possible origin of the peculiar inclination distribution of the Trojans in the scenario where Jupiter migrates inwards. We analyse different possibilities: (a) the secular evolution of an initially flat Trojan population, (b) the presence of planetary embryos among the Trojans and (c) capture of the Trojans from a pre-stirred planetesimal population in which Jupiter grows and migrates. We find that the secular evolution of the Trojans and secular perturbations from Saturn do not affect the inclination distribution of the Trojans appreciably, nor is there any significant mass depletion over the age of the Solar System. Embryos embedded in the Trojans swarms, in contrast, can stir the Trojans to their current degree of excitation and can also deplete the swarms efficiently, but it turns out that it is very difficult to get rid of all of the massive bodies in 4.5 Gyr of evolution. We propose that the disc where Jupiter's core was forming was already stirred to high inclination values by the presence of other planetary embryos competing in Jupiter's core's feeding zone. We show that the trapped Trojans preserve their high inclination through the gas phase of the protoplanetary disc and that Saturn's perturbations are more effective on highly inclined Trojans, leading to a lower capture efficiency and to a substantial depletion of the swarms during $4.5$ Gyr of evolution.}

   \keywords{minor planets, asteroids: general --
             planets and satellites: dynamical evolution and stability             
               }

   \maketitle


\section{Introduction}\label{sec:introduction}

Jupiter Trojans are a population of minor bodies in our Solar System. They share the same orbit of Jupiter with a semimajor axis of about 5.2 au and cluster in two different regions along its orbit. The leading group precedes Jupiter and librates around the L$_4$ triangular Lagrangian point; the trailing group follows Jupiter and librates around the L$_5$ triangular Lagrangian point. The number of asteroids in the leading group is larger compared to the trailing group, which is measured as an asymmetry ratio between the two swarms of $1.4\pm0.2$ for Trojans larger than 10 km in size \citep{grav11}. The Trojans possess orbits with high inclinations (up to $40^\circ$) and they are very dark objects, more similar to trans-Neptunian objects (TNOs) than asteroid belt objects. In fact, they are mainly D-type asteroids (very low albedo and relatively featureless spectra with very steep red slope) with few P-type (low albedo, featureless spectrum with reddish slope) and C-type (also low albedo, carbon-rich) asteroids \citep{barucci02,demeo14} in contrast to the main belt.

Jupiter Trojans could represent the key to understand the formation and evolution of the early Solar System. Their peculiar characteristics, such as the asymmetry ratio, the high inclination distribution and the predominance of D and P type asteroids among them, must be explained in order to really unveil the history of Jupiter and so that of the entire Solar System.

Among the most plausible hypotheses for the origin of Jupiter Trojans, there is the so-called ``chaotic capture'' \citep{morbidelli05}: any primordial Trojan is lost during the late instability of the giant planets \citep{morbidelli05,tsiganis05,gomes05} as Jupiter and Saturn cross their mutual 2:1 mean motion resonance. The swarms are then refilled with TNOs destabilised by the outward migration of Neptune. TNOs are very dark objects and the ones captured as Trojans possess also a high inclination distribution. Despite successfully matching these features, the model suffers from a low capture probability, between $10^{-6}$ and $10^{-5}$ \citep{lykawka10}, and it provides no explanation for the asymmetry ratio between the two Trojan swarms.

In the ``jump capture'' \citep{nesvorny13}, instead, the presence of a fifth giant planet in the very early Solar System is invoked. When the system becomes unstable, Jupiter has multiple close encounters with this additional planet. As a result, the semimajor axis of Jupiter ``jumps'' and radially displaces the Trojan stable regions, losing the primordial Trojans and capturing new asteroids with semimajor axis similar to its new position. At the time of the last jump of Jupiter's semimajor axis (that is when Trojans are captured) the planet vicinity was populated with trans-Neptunian objects destabilised by the outward migration of Neptune.
This model reproduces the orbital distribution of the Trojans, their dark photometric colour and it is also potentially capable of explaining their asymmetry ratio: in case the extra ice giant involved in the planet-planet scattering with Jupiter traverses one of the Trojan swarms, it can scatter captured bodies out of the stable region, depleting the swarm. However, even if the extra ice giant traverses the correct swarm, the results in \citet{nesvorny13} cannot rule out a symmetric ratio between the swarms within $1\sigma$. Also, the low capture probability, of the order of $6-8 \times 10^{-7}$, is a weakness in this model too. 

Recently, \citet{pirani19} showed that the asymmetry ratio of the Trojans could arise as a direct consequence of the early inward migration of Jupiter through the gaseous protoplanetary disc phase while it was growing to become a gas giant. In this scenario, Jupiter's core grows according to the core accretion model \citep{pollack96} boosted by pebble accretion \citep{johansen10,ormel10,lambrechts12,ida16,johansen17} and migrates inwards due to interactions with the gaseous disc \citep{ward97,lin86}, following ``growth tracks'' similar to those shown in \citet{bitsch15b}. In order for Jupiter to end its inward migration at about 5 au when the gaseous disc disperses, its core has to form in the outer Solar System. Trojans are then captured in the feeding zone of Jupiter's core, at about 20 au, among objects that naturally posses dark photometric colours like the current Jupiter Trojans, and dragged by the migrating planet to where they currently orbit. The relative drift between the planet and the Trojans induces a deformation of the horseshoe orbits of asteroids in resonance with Jupiter, leading to an excess of objects in the L$_4$ side of the horseshoe region. The mass growth of Jupiter then shrinks these orbits into tadpole orbits, originating the asymmetry. Despite this good agreement with observations, the simulations of \citet{pirani19} showed a final mass of the Trojans that is 3-4 orders of magnitude higher than the current one and an inclination distribution that is much flatter than the current one. 

The high inclination distribution of the Trojans is a long-standing problem in the models where Trojans are captured from planetesimals orbiting around Jupiter during its growth, the so-called ``local capture models'' \citep{marzari02}. Different solutions have been proposed to drive them into high-inclination orbits: the raising of inclinations by secular resonances \citep{marzari00}; a process analogous to that suggested by \citet{wetherill92} and \citet{petit01}, that is to posit the presence of the massive embryos in the Trojan swarms that excited the population by their gravity and were ejected from the stable regions by mutual perturbations; the possibility that Trojans were stirred up prior to capture by proto-Jupiter, as suggested in \citet{marzari02}.

The aim of this follow-up paper is to address the Trojan mass and inclination issues identified in \citet{pirani19}. In order to do that, we explore three different plausible ways to incline the Trojans up to $40^\circ$ and keep track of the mass depletion in each different scenario. Under the influence of Jupiter and Saturn, in this paper we test: (a) the secular evolution of an initially flat Trojan population, (b) the presence of massive planetary embryos among the Trojans and (c) a pre-stirred disc planetesimal population in which Jupiter is growing and migrating.

The paper is organised as follows: in section \ref{sec:methods} we describe the different scenarios and methodology used in our simulations; in section \ref{sec:results} we present our results. Finally, in section \ref{sec:conclusions} we summarise our results and discuss their implications.

\section{Methods}\label{sec:methods}

In our simulations, we utilised a parallelised version of the \textsc{Mercury} \textit{N}-body code \citep{chambers99} and we selected its hybrid symplectic integrator that is faster than conventional N-body algorithms by about one order of magnitude \citep{wisdom91} and particularly suitable for our simulations that involve timescales of the order of billions of years. We used a time step of 140 days that is about 1/20 of the orbital period of a particle orbiting at about 4 au \citep{duncan98}. Since we are interested in the Trojans that orbit with Jupiter at 5.2 au, this is a sufficient resolution. We modified the code so that the giant planets grow and migrate according to the growth tracks generated following the recipes in \citet{johansen17}, as will be explained in subsection \ref{sec:gt}. 

In the \textsc{Mercury} \textit{N}-body code, the planets and planetary embryos are treated as massive bodies, so they perturb and interact with all the other bodies during the integration. The other particles, called small bodies, are perturbed by the massive bodies but cannot affect each other. Since we set them as ``massless'', they also cannot perturb the massive bodies. We will refer to these particles in the text as ``massless particles'' or ``small bodies''. In our simulations we used these massless particles to populate the protoplanetary disc in which Jupiter is growing and migrating.
Our version of the code also includes aerodynamic gas drag effects and tidal gas drag effects to mimic the presence of the gas in the protoplanetary disc as in \citet{pirani19}. The growing protoplanets and planetary embryos are affected by the tidal gas drag and the massless particles are affected by the the aerodynamic gas drag until the gaseous protoplanetary disc photoevaporates at $t=3$ Myr, according to typical disc lifetimes \citep{mamajek09,williams11}. Since small bodies are set to be massless during the integrations, we assign them a radius $r_{\rm{p}}=50$ km and a density $\rho_{\rm{p}}=1.0$ g/cm$^3$ when computing the effect of the aerodynamic gas drag on the particle that is the typical size resulting from streaming instability simulations \citep{johansen14}.

A key result from \citet{pirani19} is that Jupiter Trojans are almost all captured from Jupiter's core feeding zone. Because of this, our massless particle disc extends only for about $\pm$ 2.5 au from Jupiter's core's location. The disc is then divided into annular regions of 0.5 au, populated by 10000 massless particles each. The same amount of particles in each annular region means that we adopt a surface density proportional to $r^{-1}$ for the primordial bodies component. 

\subsection{Growth tracks}\label{sec:gt}

In order to generate the ``growth tracks'' for Jupiter and Saturn that we implement in our simulations, we used the recipes in \citet{johansen17}. The disc parameters for our model are: $f_{\rm{g}}=0.2$, $f_{\rm{p}}=0.4$, $f_{\rm{pla}}=0.2$, $H/r=0.04$, $H_{\rm{p}}$/$H=0.1$, $\Delta \varv= 30$ m/s and $\textit{St} =0.1$, where $f_{\rm{g}}$, $f_{\rm{p}}$ and $f_{\rm{pla}}$ are parameterisations of the column densities (of the gas, pebbles and planetesimals, respectively) relative to the standard profiles,  $H/r$ is the disc aspect ratio, $H_{\rm{p}}/H$ is the particle midplane layer thickness ratio, $\Delta \varv$ is the sub-Keplerian speed of the gas slowed down by the radial pressure support and $\textit{St}$ is the Stokes number of the pebbles. 
\begin{figure}
\begin{center}
\includegraphics[width=\hsize]{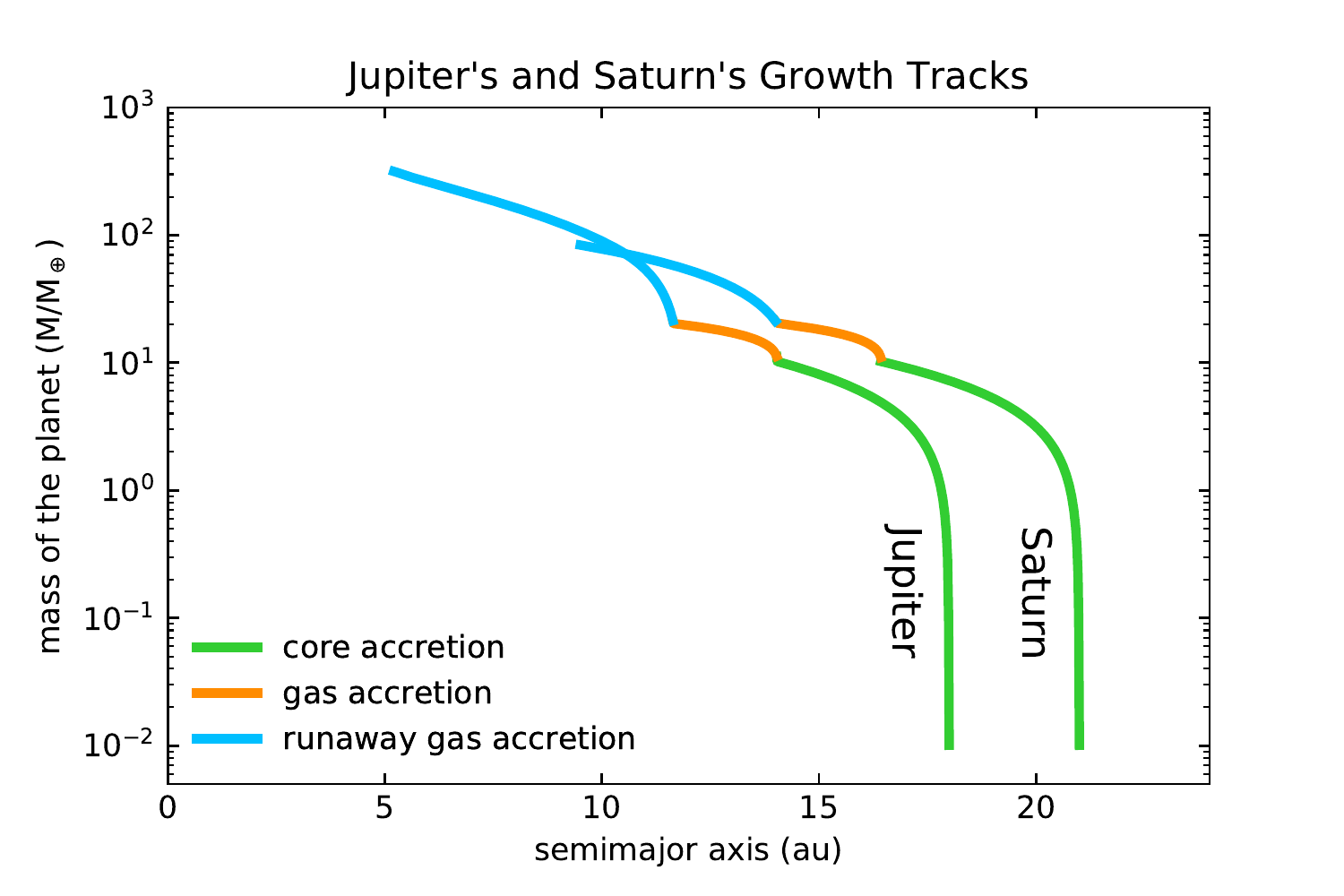}
\caption[]{Jupiter's and Saturn's growth tracks. The gas giants start with an initial mass of $10^{-2}$  M$_\oplus$ and grow to their current mass. Jupiter migrates from 18 au to its current orbit at 5.2 au and Saturn from 21 au to 9.5 au. The migration starts at $\sim$2.2 Myr and stops when the gas dissipates at 3 Myr. The green solid line corresponds to the core accretion phase; the orange solid line corresponds to the gas accretion phase and the cyan solid line corresponds to the runaway gas accretion phase.}
\label{fig:growth_tracks}
\end{center}
\end{figure}
Jupiter's and Saturn's growth tracks are shown in Figure \ref{fig:growth_tracks}. The initial mass of Jupiter's seed is $10^{-2}$ M$_\oplus$ and its final mass is about 300 M$_\oplus$. It migrates from 18 au to its current orbit at 5.2 au. The migration starts at about 2.2 Myr and stops when the gas dissipates at 3 Myr. The green solid line corresponds to the core accretion phase; the orange solid line corresponds to the gas accretion phase and the cyan solid line corresponds to the runaway gas accretion phase. Saturn's growth track is similar to Jupiter's. It starts as massive as Jupiter's seed ($10^{-2}$ M$_\oplus$) and reaches a final mass of 95 M$_\oplus$. It migrates from 21 au to 9.5 au in about 0.8 Myr. 

We do not simulate any late instability of the giant planets like the Nice Model \citep{morbidelli05,tsiganis05,gomes05} and the fifth giant planet model \citep{nesvorny11}, nor do we lock the planets in any mutual resonances. Time-scales and the time when the late instability occurs are still debated \citep{morbidelli18}, as is the time (if any) the planets spend in mean motion resonance. The amount of depletion to be attributed to the late instability is very dependent on which version we consider and we are not going to explore it in this paper. \citet{pirani19} showed that Trojans and their original asymmetry can survive a single jump of Jupiter of 0.2 au. In line with this, we will attribute a fictitious 80\% of depletion of the Trojans to the late instability when we will analyse our results. We also do not include the so-called Grand Tack \citep{walsh11} hypothesis, where Jupiter is supposed to migrate inwards in the inner Solar System, deep to about 1.5 au before Saturn (also migrating inwards) is caught in a 2:3 mean-motion resonance with it. At this point, Jupiter migration changes direction and the giant planets move outwards, which explains the mixing, the excitation, the depletion of the main asteroid belt and the low mass of Mars. Since we will not focus on the asteroid belt and since a slightly deeper migration will not alter our results on the Trojans significantly, we will proceed with the scenario in which Jupiter reaches 5.2 au when the protoplanetary disc photoevaporates at 3 Myr.

\subsection{Secular evolution of massless Trojans} \label{sec:flat}

In the first set of simulations, we generated massless particles with random eccentricities in the interval $[0,0.01]$, random inclinations in the interval $[0^\circ,0.01^\circ]$ and random semimajor axes in each $\Delta a=0.5$ au annular region. We used a flat inclination distribution for our unperturbed disc particles. In each annular region we placed $10^4$ massless particles, from 15.5 to 20.5 au for a total of $10^5$ particles. As shown in Figure \ref{fig:growth_tracks}, Jupiter's core starts at 18 au, so it is placed exactly in the middle of the particle disc. We called these two simulations \textsc{J0}, where only Jupiter is present, and \textsc{JS0}, where Saturn is also added to the system.

Jupiter grows and migrates following the growth track in Figure \ref{fig:growth_tracks} and traps a Trojan population compatible in mass and inclination distribution with the ones reported in \citet{pirani19}.
We analysed both the case with only Jupiter, where the final configuration is Jupiter at 5.2 au with a circular orbit and zero inclination, and the one with Jupiter plus Saturn, where we let the two giant planets migrate to their current orbits and artificially smoothly increased their inclination and eccentricity to their current values following the exponential laws present in \citet{pirani19} with an e-folding time of $\tau=5$ Myr. The starting inclination distribution of the Trojans is evaluated at 5 Myr, that is 2 Myr after Jupiter reaches its current semimajor axis. This is because we want to avoid to count eccentric interlopers stirred by the migration as Trojans. 
\begin{figure}
\begin{center}
\includegraphics[width=\hsize]{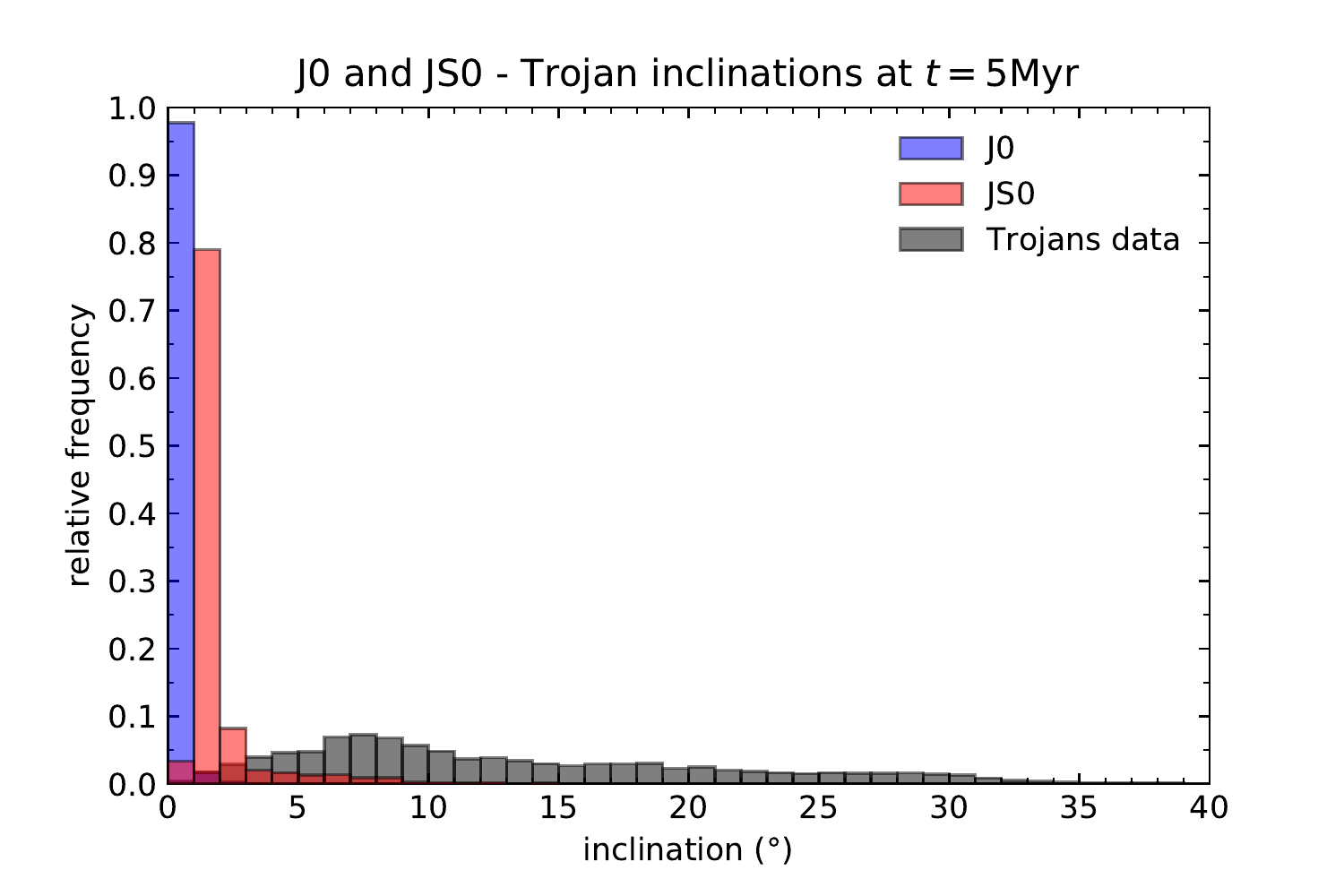}
\caption[]{Starting inclinations for \textsc{J0} (blue histogram) and \textsc{JS0} (red histogram) at 5 Myr. The grey histogram represents the current inclination distribution of Jupiter Trojans.}
\label{fig:gt}
\end{center}
\end{figure}
Figure \ref{fig:gt} shows the initial Trojan inclinations for simulations \textsc{J0} (in blue) and \textsc{JS0} (in red). The grey histogram represents the current inclination distribution of Jupiter Trojans. As expected from the results in \citet{pirani19}, the distributions remain very flat after the large-scale migration and growth of the giant planet. This is because inward migration and mass growth of Jupiter do not affect Trojan inclinations, due to the quasi-invariance of the inclination under mass growth and migration \citep{fleming00}.

\begin{table}
\caption{Initial Trojan populations in \textsc{J0} and \textsc{JS0} simulations at $t=5$ Myr.}  
\label{table:massless}                
\centering          
\begin{tabular}{c |c |c |c }     
\hline\hline       
Simulation&Initial number&Mass & Initial asymmetry\\
name& of Trojans&(M$_\oplus$)&  ratio ($N_{\rm{L4}}/N_{\rm{L5}}$)\\
\hline                    
 J0 & 2600 & 0.13 &1.60 \\
 JS0 & 2208 & 0.11 &1.48 \\     
\hline           
\end{tabular}
\end{table}

The starting number of Trojans, mass and asymmetry ratio for simulations \textsc{J0} and \textsc{JS0} is summarised in Table \ref{table:massless}. In order to estimate the initial mass, we will consider the Minimum Mass Solar Nebula (MMSN) \citep{weidenschilling77,hayashi81}, that predicts approximately 1 M$_\oplus$ of mass in each astronomical unit annular region. Since we have 10000 particles in each 0.5 annular region, a massless particle in our simulations represents $5\times10^{-5}$ M$_\oplus$. We are aware that the MMSN model cannot be too accurate since the planets migrate through the disc during their formation, but our main purpose is to assess the mass depletion as a relative value to the initial mass, and not the absolute value, in the different scenarios.

\subsection{Embryos embedded in the trojan swarms} \label{sec:sfd}

In this second scenario, we take advantage of the Trojan population trapped in the ``flat particle disc'' cases discussed in subsection \ref{sec:flat}.
Of the resulting initial Jupiter Trojans of the previous set of simulations, at $t=5$ Myr we substituted part of them in the \textsc{Mercury} \textit{N}-body code as massive bodies according to the size frequency distribution obtained from the planetesimal formation simulations of \citet{schafer17} and restarted a second separate set of simulations.
The size frequency distribution is consistent with an exponentially tapered power law with an exponential cutoff at the high-radius end:
\begin{equation}
\frac{N_>(R)}{N_{\rm{tot}}}=\left(\frac{R}{R_{\rm{min}}}\right)^{-3\rm{\alpha}}\exp\left[\left(\frac{R_{\rm{min}}}{R_{\rm{exp}}}\right)^{3\rm{\beta}}-\left(\frac{R}{R_{\rm{exp}}}\right)^{3\rm{\beta}}\right].
\end{equation}
Here $N_>(R)$ is the number of Trojans with radius greater than $R$, where $R$ is the radius of each Trojan. $N_{\rm{tot}}$ is their total number, $R_{\rm{min}}$ is the minimum Trojan radius and $R_{\rm{exp}}$ is the exponential cutoff radius. $\rm{\alpha}$ is the power law exponent of the exponentially tapered power law and $\rm{\beta}$ is the steepness of the exponential cutoff.
We set $N_{\rm{tot}}$ in order to represent roughly the total initial mass of the Trojans estimated with the MMSN model and reported in Table \ref{table:massless}, 
$\rm{\alpha}=0.6$ and $\rm{\beta}=0.35$. 

\subsubsection{More and less massive embryos}

As regards $R_{\rm{min}}$ and $R_{\rm{exp}}$, we simulated two different distributions:
\begin{enumerate}[(i)]
\item $R_{\rm{min}}=100$ km and $R_{\rm{exp}}=250$ km, where the most massive Trojan of the distribution has a mass of $2\times10^{-3}$ M$_\oplus$ (Pluto-like bodies).
\item $R_{\rm{min}}=10$ km and $R_{\rm{exp}}=100$ km, consistent with characteristic radii inferred for TNOs \citep{abod18}, where the most massive Trojan has a mass of $3\times 10^{-4}$ M$_\oplus$ (Ceres-like bodies).
\end{enumerate}
\begin{figure}
\begin{center}
\includegraphics[width=\hsize]{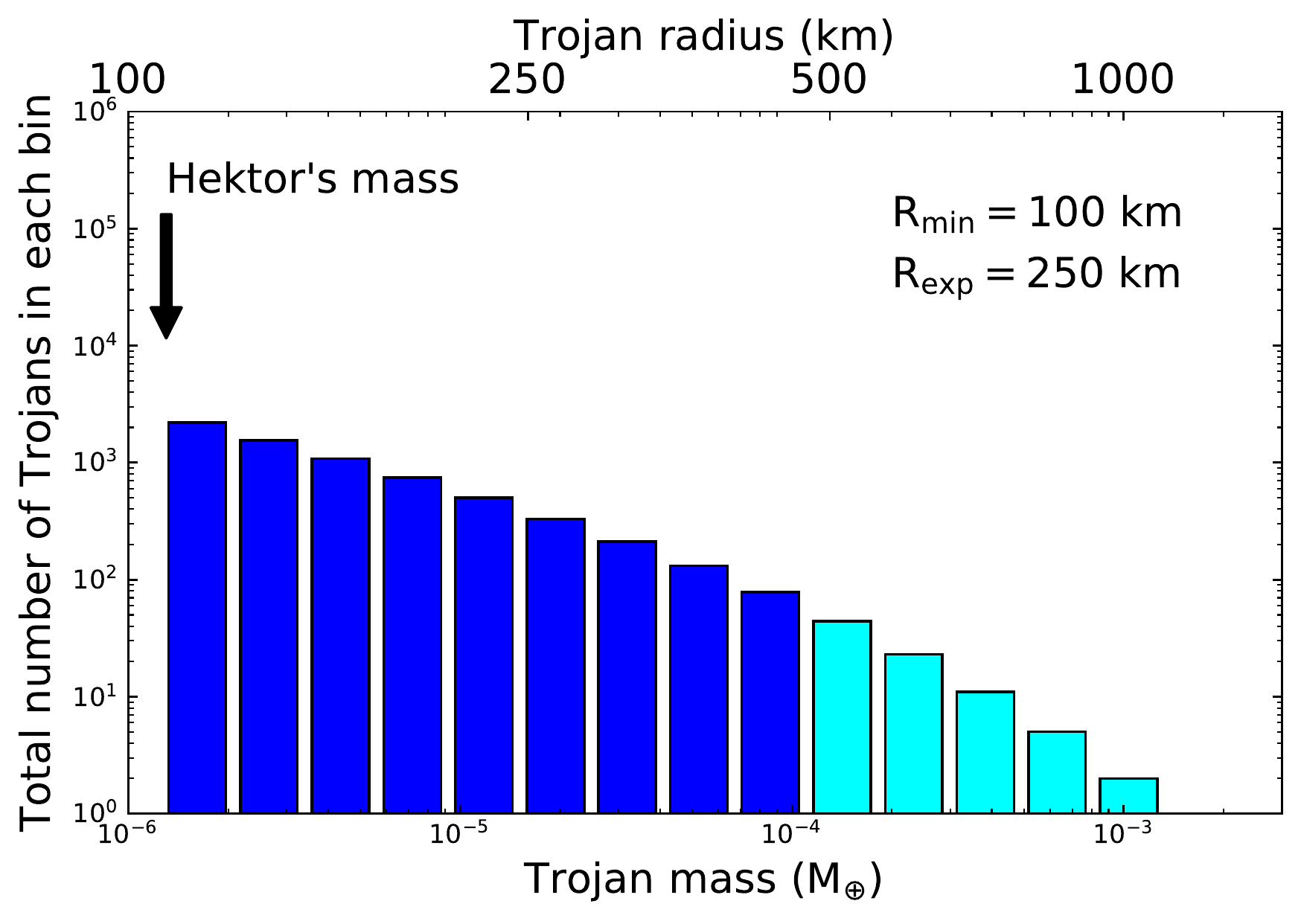}
\includegraphics[width=\hsize]{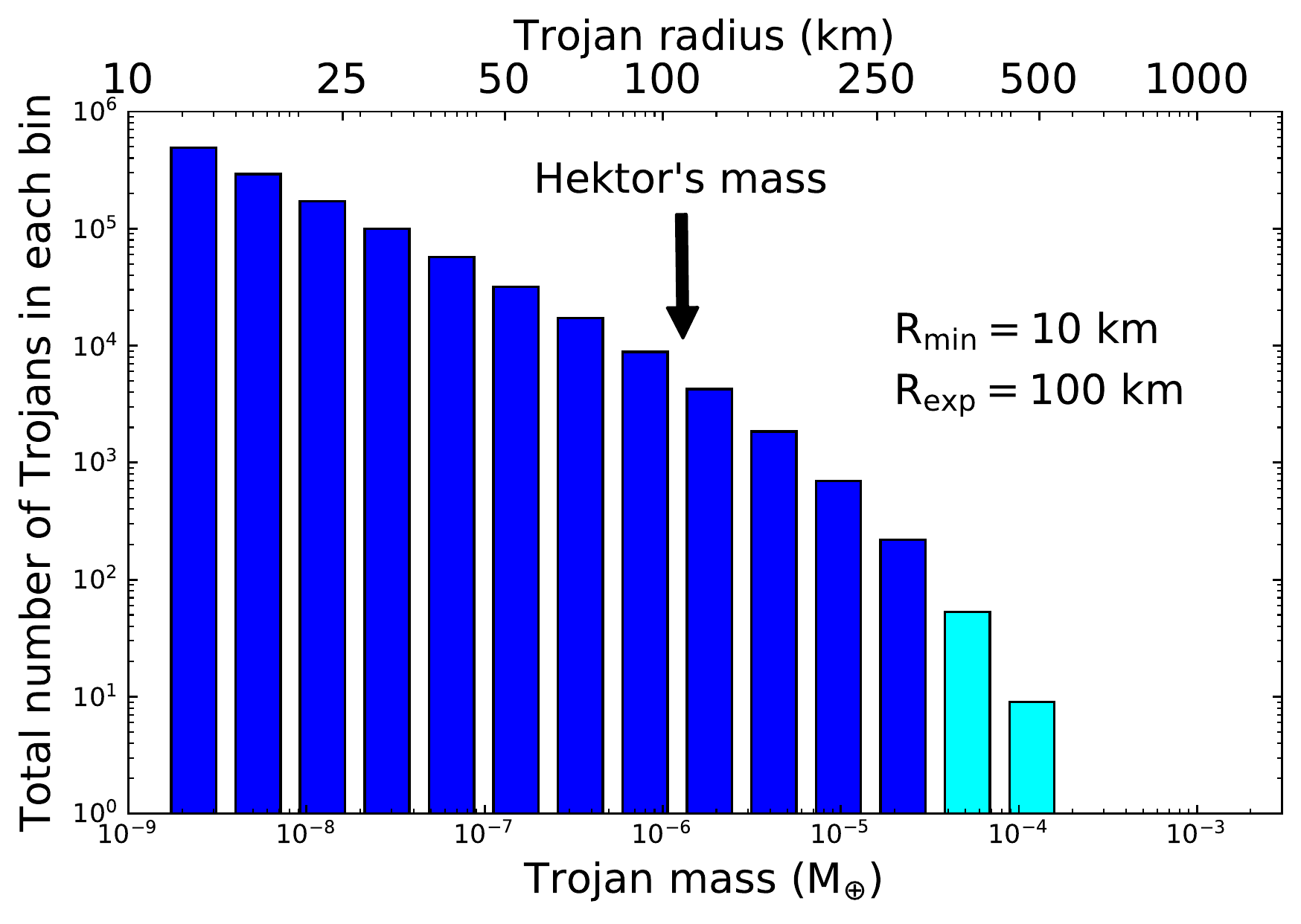}
\caption[]{Size frequency distributions of the Jupiter Trojans in the case of $R_{\rm{min}}=100$ km and $R_{\rm{exp}}=250$ km (top histogram) and in case of $R_{\rm{min}}=10$ km and $R_{\rm{exp}}=100$ km (bottom histogram). The histograms show the number of Trojans in mass bins (bottom x-axis). In the upper x-axes is shown the correspondent radius bins (we assumed a density of the Trojans of 1.5 g/cm$^3$). In cyan we highlighted the part of the distributions that we substituted into the Trojan swarms as massive bodies. The black arrow indicates the mass of the asteroid (624) Hektor, that is the most massive Jupiter Trojan.}
\label{fig:sfd}
\end{center}
\end{figure}
The size frequency distributions in the two different cases are shown in Figure \ref{fig:sfd}. The top histogram represents the case with $R_{\rm{min}}=100$ km and $R_{\rm{exp}}=250$ km and the bottom histogram represents the case with $R_{\rm{min}}=10$ km and $R_{\rm{exp}}=100$ km. Since the total mass is the same, but the range of the object sizes of the distributions is different, the total number of asteroids in the two cases differs. The black arrow in the figures indicates the mass of the asteroid (624) Hektor of about $7.9 \times 10^{18}$ kg \citep{marchis14}, the most massive Jupiter Trojan. For reasons of computational time, we substituted just the most massive part of the distribution into the Trojan population and we indicated it in cyan; in blue we show the rest of the size frequency distribution that we did not consider. In the case of $R_{\rm{min}}=100$ km and $R_{\rm{exp}}=250$ km, we substituted 86 bodies; in the case of $R_{\rm{min}}=10$ km and $R_{\rm{exp}}=100$ km, we substituted 63 bodies. These numbers depend on the logarithmic bins we used in between the intervals $R_{\rm{min}}$ and $R_{\rm{max}}$, where $R_{\rm{max}}$ is the object's maximum size of the distribution.

For this second set of simulations, we also considered the case of Jupiter migrating alone and the case in which also Saturn is migrating together with it. We named the simulations where only Jupiter is migrating as \textsc{J100} and \textsc{J250} with $R_{\rm{exp}}=100$ km and $R_{\rm{exp}}=250$ km values for the exponential cutoff radius, respectively; \textsc{JS100} and \textsc{JS250} where also Saturn is added to the system, with $R_{\rm{exp}}=100$ km and $R_{\rm{exp}}=250$ km values for the exponential cutoff radius, respectively. 
We substituted the particles in a completely arbitrary way: we substituted the first 86 (or 63) first massless particles in the input file with massive embryos without knowing if the substituted particle belongs to the L$_4$ or L$_5$ swarm, since in the file they are ranked by their initial position in the disc and hence in a random order. Simulations will stop at $t=4.5$ Gyr and we will assess the depletion history of the swarms, the fate of the embryos embedded in the swarms and if the asymmetry is sensitive to the presence of embryos.

\subsubsection{The loss of the embryos from the Trojan swarms}

The previous kind of simulations, with almost hundred of massive bodies, are very computationally expensive even when we only try to include just a small part of the distribution as massive bodies. We decided to run an additional subset of 10 simulations in which we substituted just the 10 most massive bodies from the distribution with $R_{\rm{min}}=10$ km and $R_{\rm{exp}}=100$ km. We named these simulations \textsc{10embryos}. We considered only the case where both Jupiter and Saturn are migrating. The aim of these simulations is to try to understand if it is easy to lose the massive Trojans from both the two swarms. Also in this case, we substituted arbitrarily the massive bodies, without caring if they belongs to L$_4$ or L$_5$ swarm and the starting distribution of the massive embryos within the two swarms are reported in Table \ref{table:10embryos} together with the results after $t=4.5$ Gyr. 

\subsection{Pre-stirred planetesimal disc}\label{sec:prestirred_method}

In the third and final scenario, we started with a disc that was already pre-stirred prior to the capture of Trojans by proto-Jupiter \citep{kokubo00}. For the initial inclination and eccentricity distributions we used two different ones: in the first case we are using the current inclination and eccentricity distribution of the main asteroid belt that is thought to have been depleted and excited by the presence of planetary embryos \citep{wetherill92,petit98,chambers01,petit01,bottke05,obrien07}. As modelled in \citet{minton10}, the initial eccentricity distribution of the disc particles can be modelled as a Gaussian with the peak at $\mu_e=0.15$, a standard deviation $\sigma_e=0.07$ and a lower cutoff at zero. The initial inclination distribution is a Gaussian with the peak at $\mu_i=8.5^\circ$, a standard deviation $\sigma_i=7^\circ$ and a lower cutoff at $0^\circ$. 
\begin{figure}
\begin{center}
\includegraphics[width=\hsize]{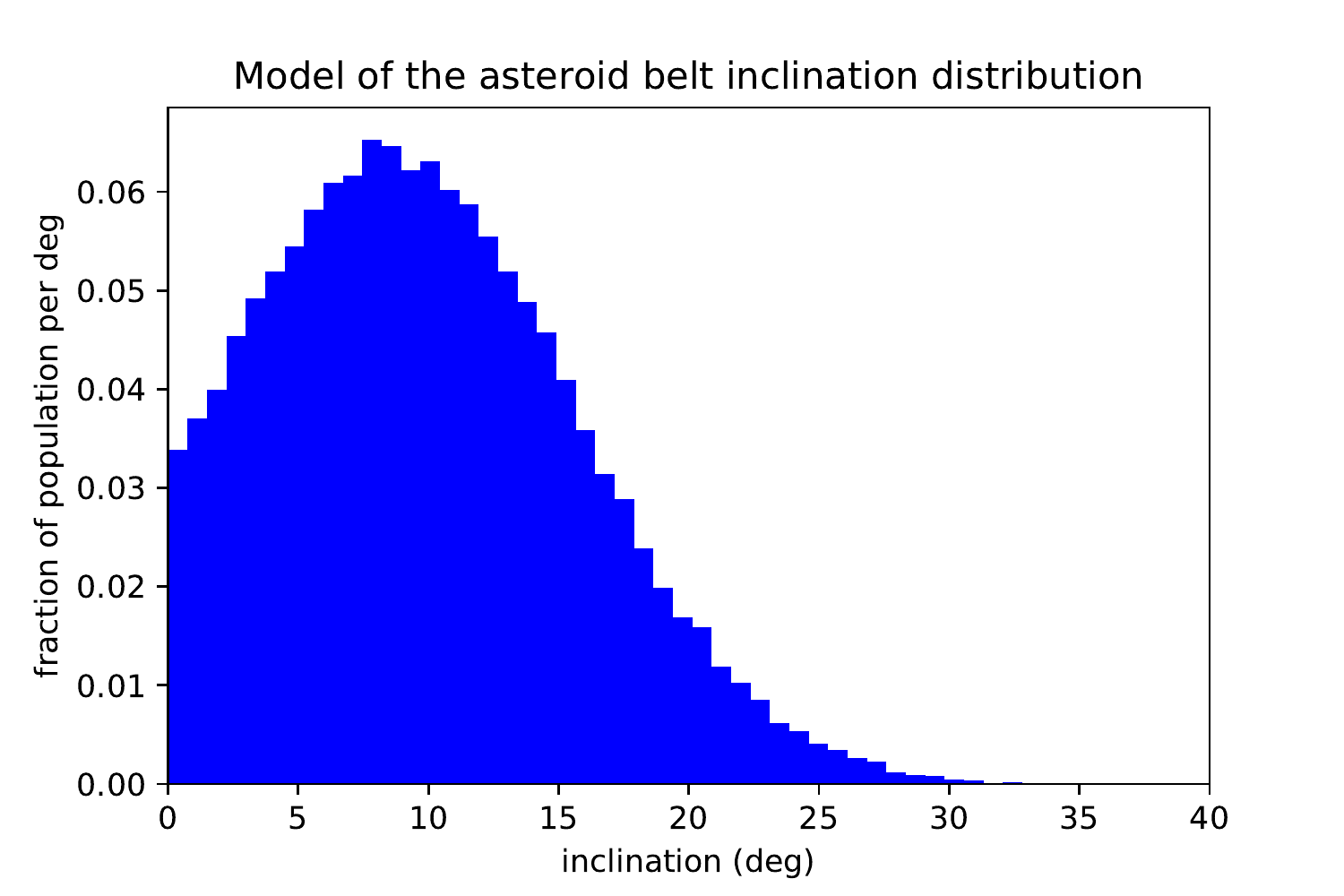}
\includegraphics[width=\hsize]{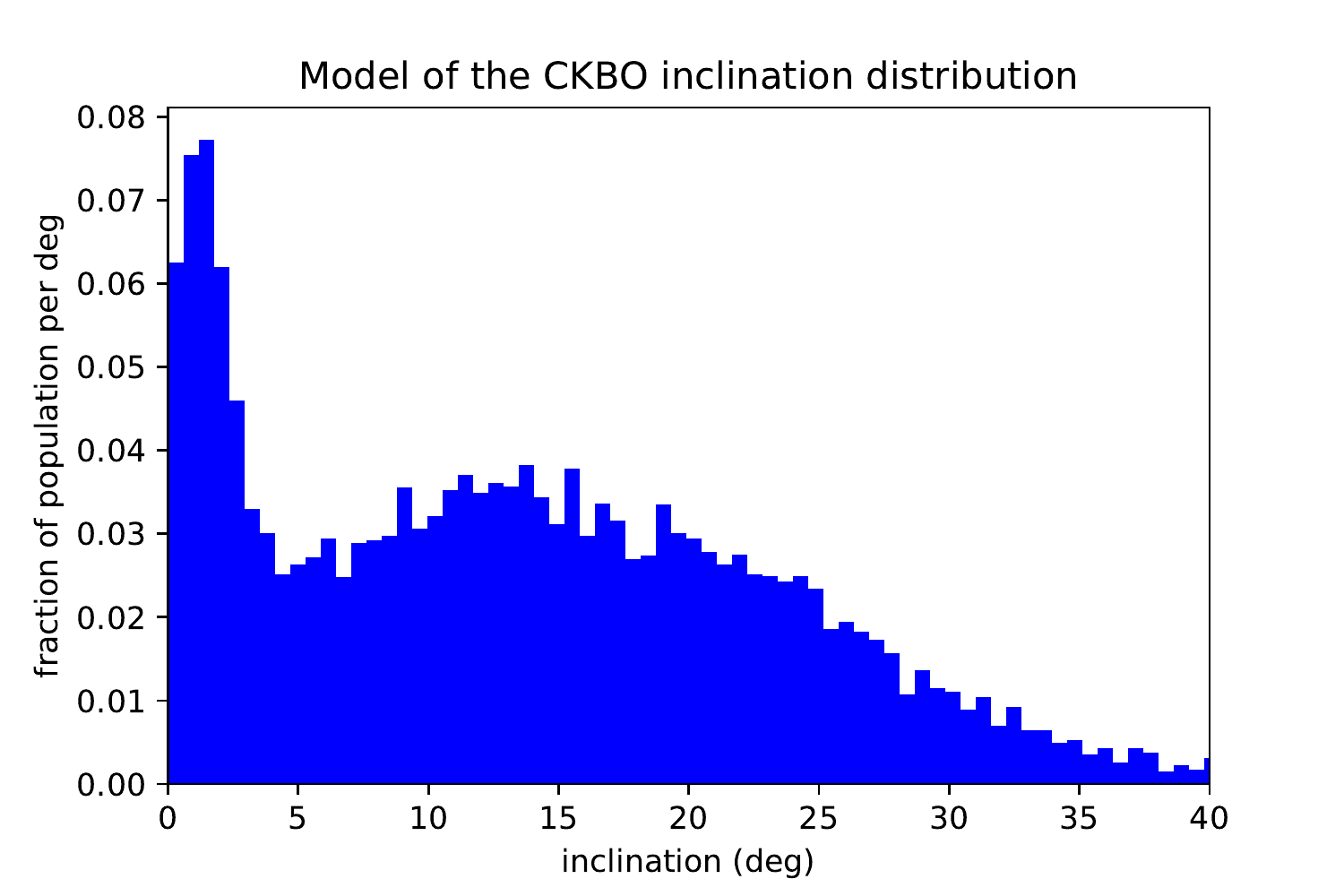}
\caption[]{In the \textsc{Jprestirred\_ab} and \textsc{JSprestirred\_ab} simulations we used an inclination distribution model similar to the asteroid belt one (top histogram). In the \textsc{Jprestirred\_ckbo} and \textsc{JSprestirred\_ckbo} simulations we used an inclination distribution model similar to the the CKBOs one (bottom histogram).}
\label{fig:incl_distributions}
\end{center}
\end{figure}
We named the simulation with only Jupiter \textsc{Jprestirred\_ab} and the one including also Saturn \textsc{JSprestirred\_ab}, where ``ab'' stands for asteroid belt. The initial inclination distribution of the particles in this case is shown in Figure \ref{fig:incl_distributions}, top histogram.
In the second case, we use the current inclination and eccentricity distribution of the Classical Kuiper Belt Objects (CKBOs), that is Hot Classicals (HCs) plus Cold Classicals (CCs) ones. We modelled them as in \citet{volk11} to a sum of two Gaussians. The initial inclination distribution is shown in Figure \ref{fig:incl_distributions}, bottom histogram. 
We named these simulations \textsc{Jprestirred\_ckbo} in the case where only Jupiter migrates and \textsc{JSprestirred\_ckbo} in the case where also Saturn is included. ``ckbo'' stands for Classical Kuiper Belt Objects.

As for the first two scenarios, we will run the simulations for $t=4.5$ Gyr in order to assess the capture efficiency of the Trojans compared to the flat disc case, the depletion history of the swarms and the asymmetry evolution, if any.


\section{Results}\label{sec:results}

\subsection{Secular evolution of Jupiter Trojans with a flat inclination distribution (\textsc{J0} and \textsc{JS0})}


\begin{table}
\caption{Evolution of the number of Trojans, their total mass and the asymmetry ratio for simulations J0 and JS0.}             
\label{table:Jmassless}      
\centering          
\begin{tabular}{c |c |c |c |c}     
\hline\hline       
Time&Trojans&Mass&Trojan&Asymmetry\\
(Myr)&left&left (M$_\oplus$)&depletion (\%)&ratio ($N_{\rm{L4}}/N_{\rm{L5}}$)\\

\multicolumn{5}{c}{\textsc{J0}}\\
\hline                    
 5&2600&0.13&0.0&1.60\\
 1000&2592&0.13&0.3&1.60\\
 4500&2592&0.13&0.3&1.60\\
  
\multicolumn{5}{c}{\textsc{JS0}}\\
\hline
 5&2208&0.11&0.0&1.48\\
 1000&1581&0.08&28.4&1.57\\
 4500&1345&0.07& 39.0& 1.61\\
\hline                 
\end{tabular}
\end{table}


Results of \textsc{J0} and \textsc{JS0} are summarised in Table \ref{table:Jmassless}. In the \textsc{J0} simulation, Jupiter Trojans are very stable over $t=4.5$ Gyr of evolution if only Jupiter is present in the system. This means that there is no mass depletion over the history of the Solar System and that the original asymmetry ratio between the number of Trojans in L$_4$ and L$_5$ is also preserved. 
While considering the scenario that includes both Jupiter and Saturn, we notice that interactions between the two planets during the migration led to a lower capture efficiency. Indeed, we capture 15\% less Trojans in the \textsc{JS0} simulation (2208 captured Trojans in \textsc{JS0} against 2600 in \textsc{J0}). Moreover, when the planets stop the migration, the depletion of Jupiter Trojan swarms keep going as shown in Table \ref{table:Jmassless}, fourth column. The depletion is of the order of 40\% over 4.5 Gyr of evolution.
\begin{figure}
\begin{center}
\includegraphics[width=\hsize]{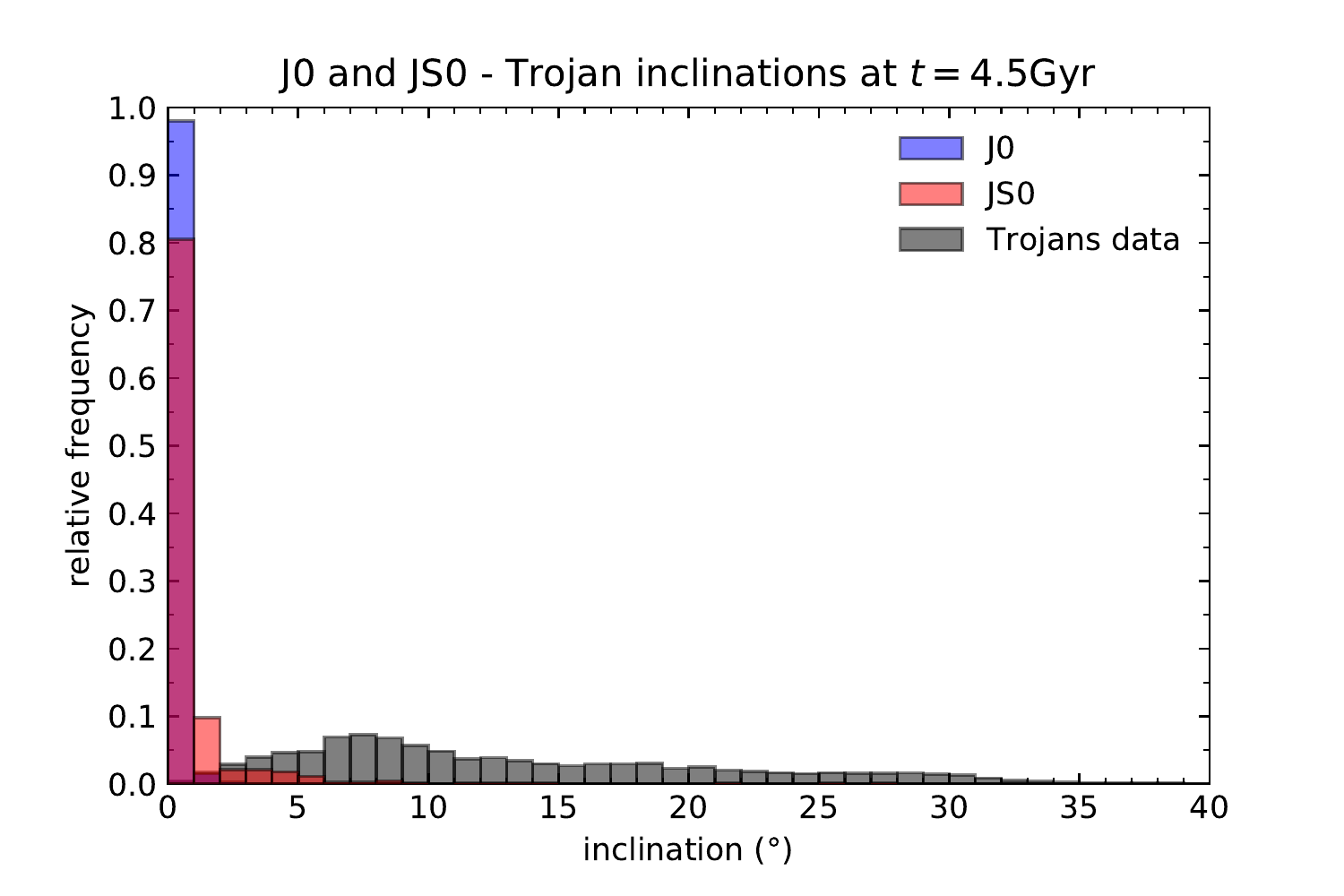}
\caption[]{Inclination distribution of the Jupiter Trojans at $t=4.5$ Gyr for the case of just Jupiter migrating (\textsc{J0}) in blue and for the case of Jupiter and Saturn both migrating (\textsc{JS0}) in red. In these simulations all the Trojans are massless. In grey the current observed inclination distribution of the Trojans.}
\label{fig:massless4500}
\end{center}
\end{figure}
As regards the final orbital parameters of the Trojans, the inclination distributions after 4.5 Gyr of evolution are shown in Figure \ref{fig:massless4500}: the Trojan inclination distribution for the simulation with only Jupiter (\textsc{J0}) is reported in blue; the Trojan inclination distribution for the simulation with Jupiter and Saturn (\textsc{JS0}) is reported in red; in grey, the current inclinations of the Jupiter Trojans is displayed. 
In both scenarios, with and without Saturn, the inclination distribution of the Jupiter Trojans remains very flat through the $t=4.5$ Gyr we integrated. These results are not in agreement with the current Trojan inclinations that are up to $40^\circ$. 


\subsection{Simulations with embryos}

\subsubsection{\textsc{J250} and \textsc{J100}}

\begin{table}
\caption{Evolution of the number of massless Trojans, planetary embryos and asymmetry ratio for simulations \textsc{J250} and \textsc{J100}.}             
\label{table:J250}      
\centering          
\begin{tabular}{c |c |c |c | c}     
\hline\hline       
Time&Total Trojans&Embryos&Depletion&Asymmetry\\
(Myr)&left&left&(\%)&ratio ($N_{\rm{L4}}/N_{\rm{L5}}$)\\                   
\multicolumn{5}{c}{\textsc{J250}}\\
\hline 
 5&2600&86&0.0&1.60\\
 100&959&4&63.1&0.81\\
 500&390&2&85.0&0.40\\
 1000&239&2&90.8&0.44\\
 2000&130&2&95.0&0.24\\
 4500&52&2&98.0&0.18\\
\multicolumn{5}{c}{\textsc{J100}}\\
\hline 
 5&2600&63&0.0&1.60\\
 100&2267&10&12.8&1.53\\
 500&1677&3&35.5&1.47\\
 1000&1303&3&49.9&1.55\\
 2000&944&2&63.7&1.76\\
 4500&528&2&79.7&1.66\\
\hline                 
\end{tabular}
\end{table}
In Table \ref{table:J250} we show the depletion history of the Trojans in simulations \textsc{J250} and \textsc{J100} during $t=4.5$ Gyr. In the first column we report the time, in the second column the total amount of massive and massless bodies left as Trojans at that time, in the third column the massive bodies left in the Trojan swarms and in the fourth column the percentage of depletion of the Trojans. 
The first important results we can infer from simulations is that embryos are very effective in depleting the Trojan swarm, even without the presence of Saturn. We obtained a depletion of 98.0 \% of the initial mass of the Trojan swarms in simulation \textsc{J250} and 79.7\% in \textsc{J100}. The issue with the embryo scenario is getting rid of the massive embryos since nowadays we do not observe any massive asteroid larger than about 200 km in diameter in the Trojan swarms: as anticipated, the largest Trojan is (624) Hektor with a mean diameter of $250 \pm 26$ km \citep{marchis14}. Indeed, even though we lost almost all the massive bodies substituted in the swarms, two of them survived as Trojans in both simulations, one in each swarm. 

Another important parameter to evaluate is the asymmetry ratio between the two swarms. In Table \ref{table:J250}, in the last column, we show the asymmetry ratio evolution. As we can see, in simulation \textsc{J250} the initial asymmetry ratio starts with a value consistent with the current observed asymmetry ratio of the Trojans, then decreases in time and eventually is reversed. This is because we completely randomised which and how many massive embryos end up in each swarm. It turned out that the leading swarm hosted the 4 most massive embryos and they depleted the leading swarm in time more effectively than the trailing swarm, reversing the asymmetry. In simulation \textsc{J100}, the asymmetry ratio remains more or less constant during 4.5 Gyr of evolution, even though 5 out of 8 of the most massive bodies ended up in the leading swarm, meaning that probably the embryos are not massive enough to affect the original asymmetry. 
\begin{figure}
\begin{center}
\includegraphics[width=\hsize]{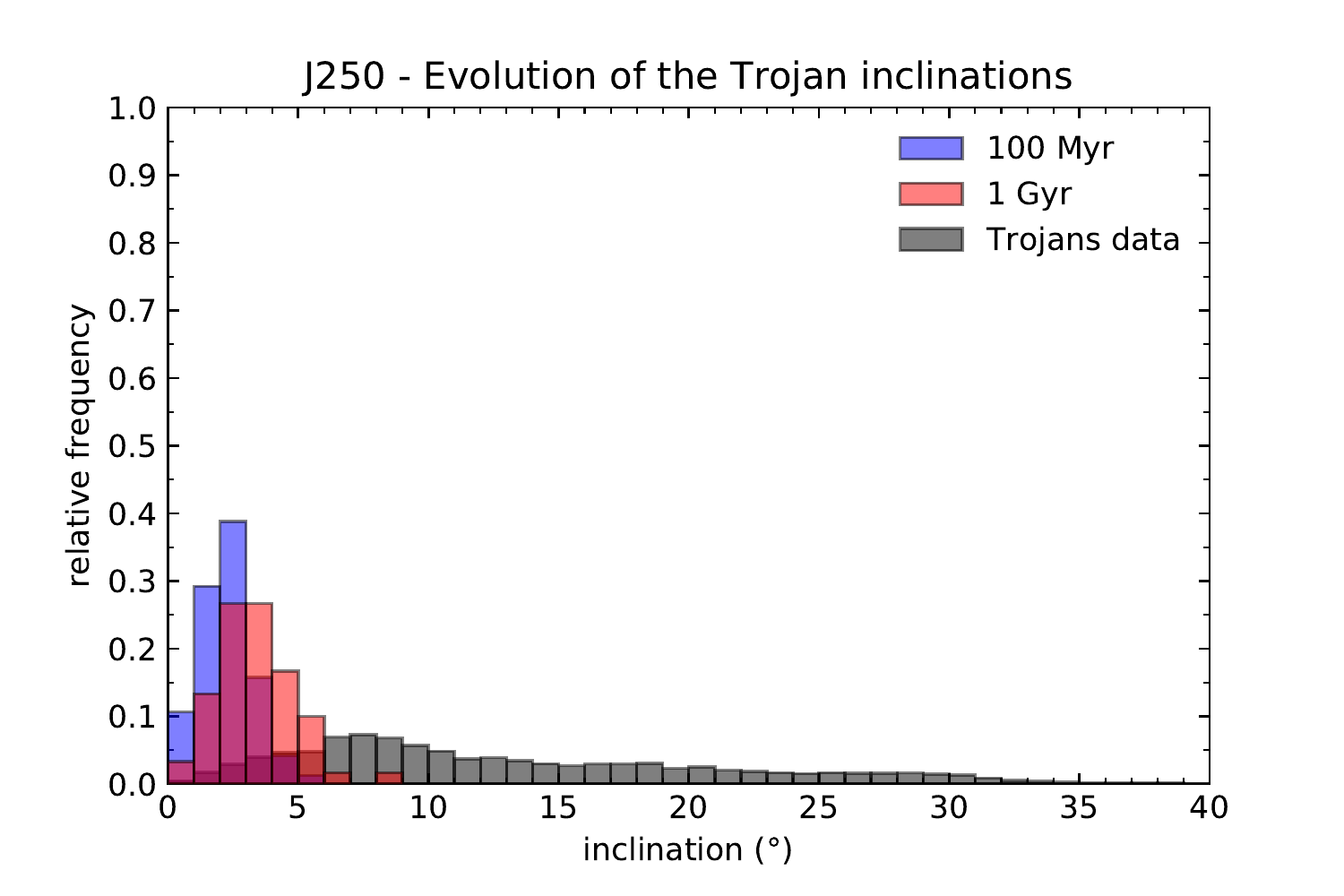}
\includegraphics[width=\hsize]{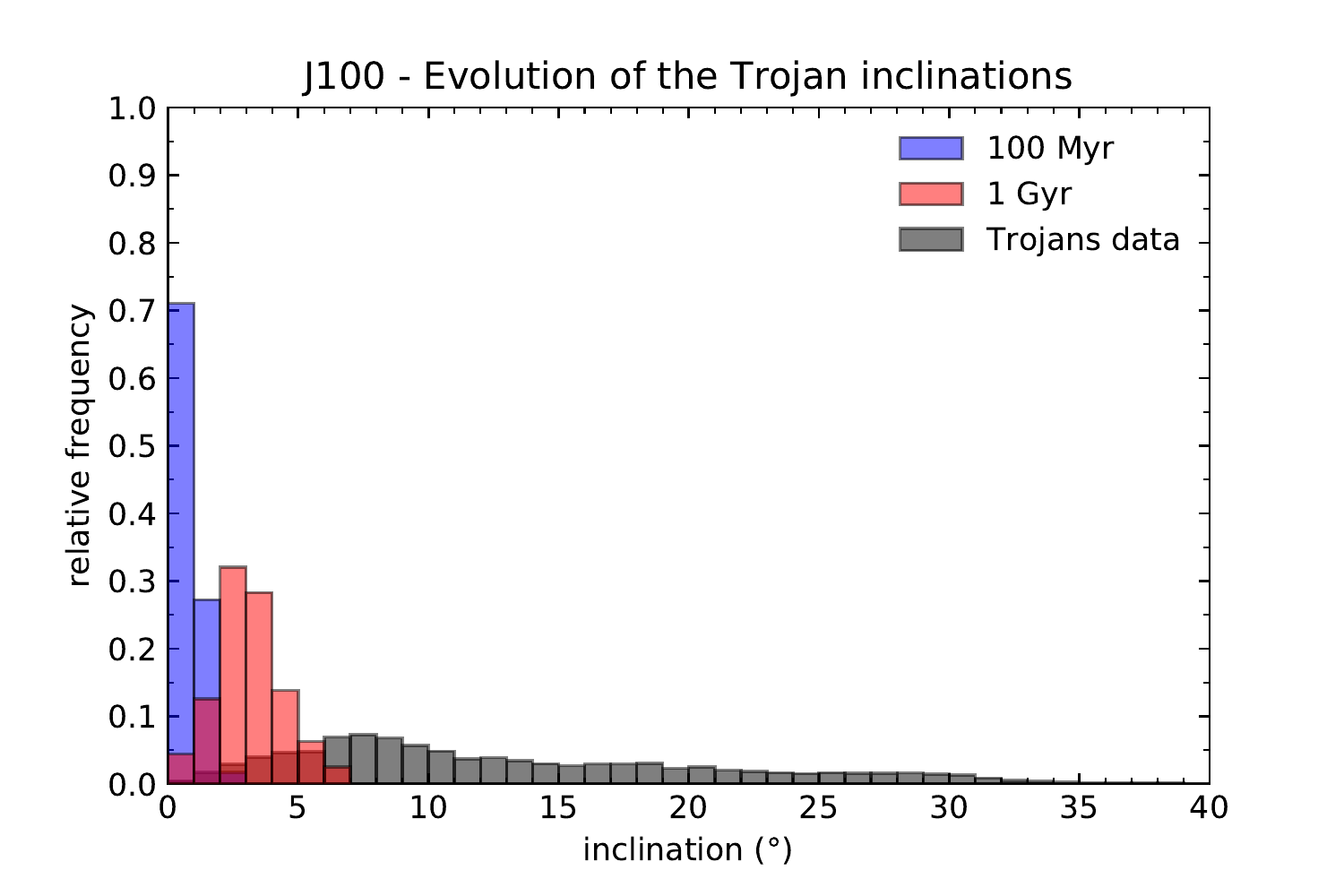}
\caption[]{Inclination distribution of the Jupiter Trojans at $t=100$ Myr (in blue) and at $t=1$ Gyr (in red). In the top histogram, the Trojan inclinations resulting from simulation J250. In the bottom histogram, the Trojan inclinations resulting from simulation J100. In grey the current observed inclination distribution of the Trojans.}
\label{fig:J250i}
\end{center}
\end{figure}

As regards the final inclination distribution of the Jupiter Trojans in simulation \textsc{J250}, it does not match with observations. Figure \ref{fig:J250i} (top histogram) shows the Trojan inclinations at 100 Myr (in blue) and at 1 Gyr (in red). Plots at 2 Gyr and 4.5 Gyr are not shown since we are left with too few Trojans to obtain a significant distribution. As we can see, Trojans acquire an inclination ranging from $0^\circ$ to $5^\circ$ that is still too small compared to the current distribution, but it is not completely flat as in the J0 and JS0 simulations.
If we analyse the inclination distribution of Jupiter Trojans in simulation \textsc{J100} (Figure \ref{fig:J250i}, bottom histogram), we report that less massive embryos can stir the Trojan inclinations as much as we obtained with Pluto sized embryos in simulation \textsc{J250} over $t=4.5$ Gyr. Again, this is not enough compared to the current inclination distribution of the Jupiter Trojans.

\subsubsection{\textsc{JS250} and \textsc{JS100}}


\begin{table}
\caption{Evolution of the number of massless Trojans, planetary embryos and asymmetry ratio for simulations \textsc{JS250} and \textsc{JS100}.}             
\label{table:JS250}      
\centering          
\begin{tabular}{c |c |c |c |c }     
\hline\hline       
Time&Trojans&Embryos&Depletion&Asymmetry\\
(Myr)&left&left&(\%)&ratio ($N_{\rm{L4}}/N_{\rm{L5}}$)\\
\multicolumn{5}{c}{\textsc{JS250}}\\
\hline                    
 0&2208&86&0.0&1.48\\
 100&1409&48&36.2&1.41\\
 500&404&20&81.7&1.37\\
 1000&159&10&92.8&0.78\\
 2000&55&5&97.5&0.46\\
 4500&9&3&99.6&0.56\\
\multicolumn{5}{c}{\textsc{JS100}}\\
\hline
  0&2208&63&0.0&1.48\\
 100&1760&38&20.3&1.49\\
 500&1325&26&40.0&1.56\\
 1000&994&20&55.0&1.68\\
 2000&649&14&70.6&1.75\\
 4500&347&8&84.3&1.65\\
\hline                
\end{tabular}
\end{table}
The depletion history of the Trojans and the number of Trojans and embryos left after $t=4.5$ Gyr in simulations \textsc{JS250} and \textsc{JS100} are shown in Table \ref{table:JS250}. 
As in simulation \textsc{J250}, we notice that more massive embryos are very effective in depleting the swarms. The depletion is of 99.6\% in simulation \textsc{JS250}. In this case we also end up with massive Trojan embryos surviving in the swarms after $t=4.5$ Gyr: this time three survive. This highlights again the fact that it is not easy to deplete the swarm from a starting massive population. In simulation \textsc{JS100}, instead, depletion is not as effective as when we used more massive Trojans. The depletion is 84.3\% of the initial Trojans and the surviving massive Trojans are 8.  
Analysing the asymmetry ratio, in simulation \textsc{JS250} it is again reversed due to the same reason as in simulation \textsc{J250}: the leading swarm hosted the more massive Trojans. In simulation \textsc{JS100} the asymmetry remains more or less of the same order as in \textsc{J100}, confirming that less massive embryos cannot influence it too much.
\begin{figure}
\begin{center}
\includegraphics[width=\hsize]{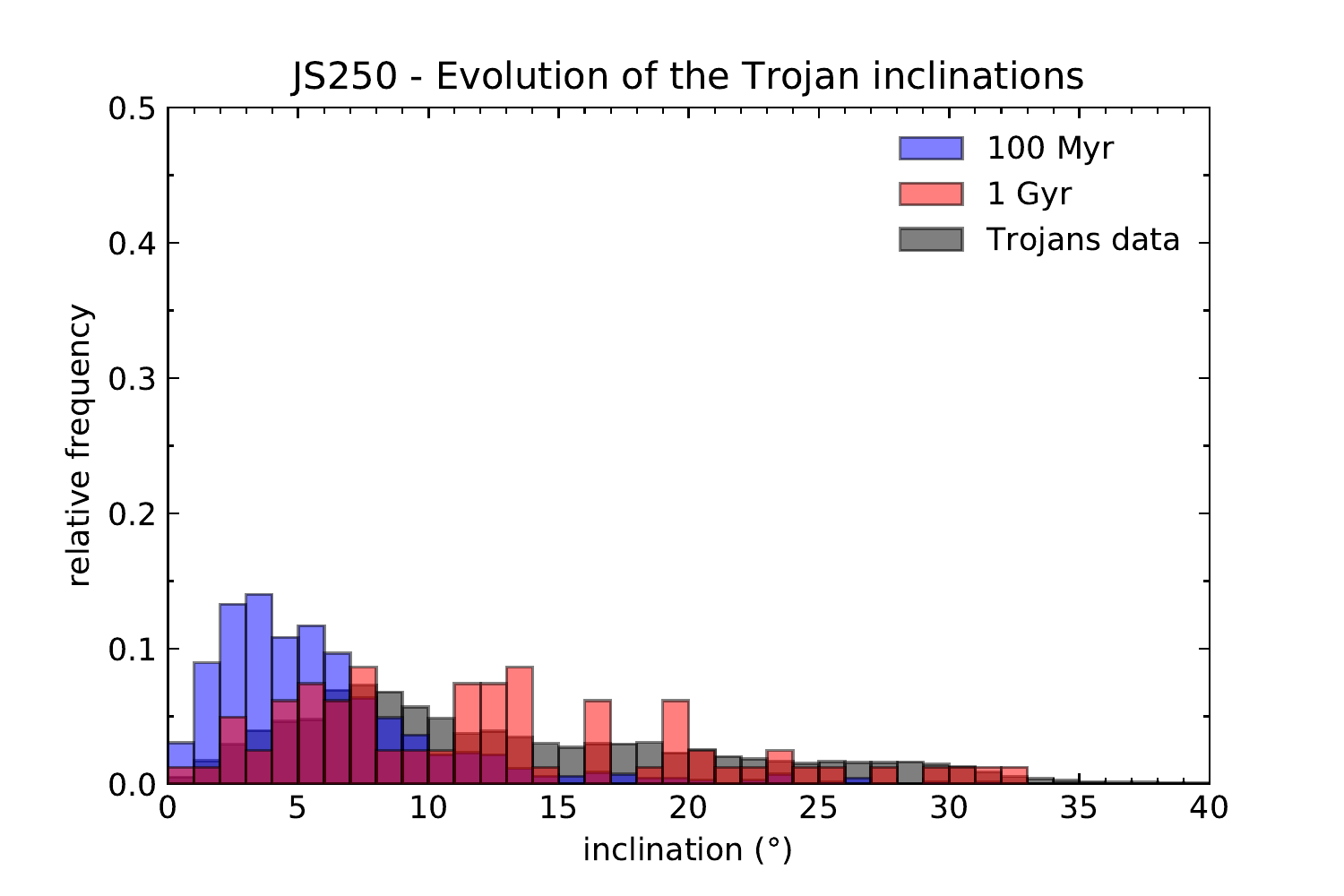}
\includegraphics[width=\hsize]{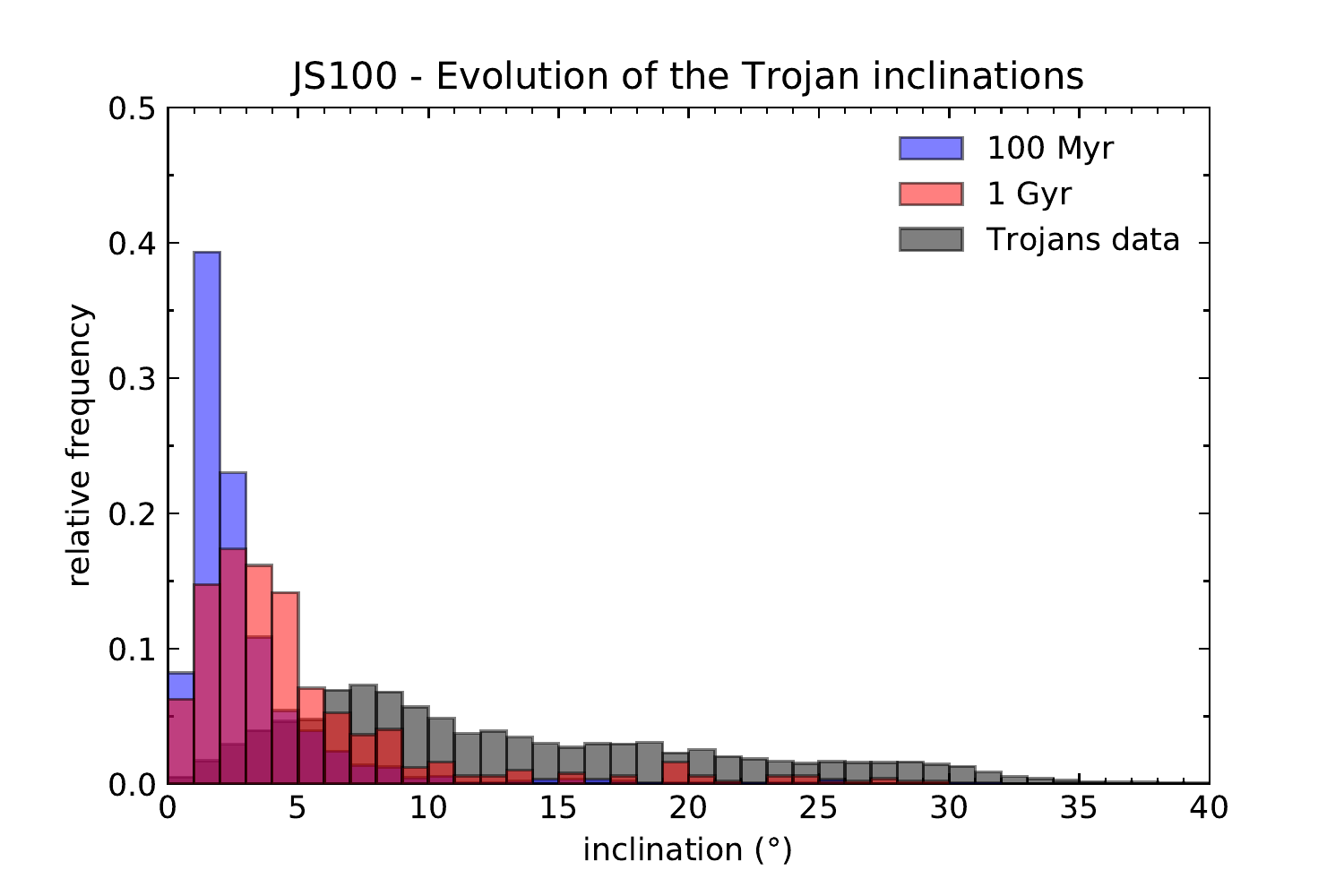}
\caption[]{Inclination distribution of the Jupiter Trojans at $t=100$ Myr (in blue) and $t=500$ Myr (in red). In the top histogram, the Trojan inclinations resulting from simulation \textsc{JS250}. In the bottom histogram, the Trojan inclinations resulting from simulation \textsc{JS100}. In grey the current observed inclination distribution of the Trojans.}
\label{fig:JS250i}
\end{center}
\end{figure}
In Figure \ref{fig:JS250i} we show the inclination distribution of the Trojans at different times during the integration: at 100 Myr (in blue) and at 1 Gyr (in red). In grey we show the observed inclination distribution of the Trojans. The top histogram shows the resulting Trojan inclinations of simulation \textsc{JS250} that are stirred by the presence of the embryos and eventually reach the current observed values. In the bottom histogram, we show the inclination distribution of the Trojans in the \textsc{JS100} simulation, where the embryos are able to stir up the distribution over the right range, but the average value is too low.

\textsc{JS250} and \textsc{JS100} simulations have slightly higher Trojan depletion than \textsc{J250} and \textsc{J100} simulations. Since this is a trend we already reported in the cases where no massive embryos where involved (that is simulations \textsc{J0} and \textsc{JS0}) we believe that this extra depletion is to be attributed to the presence of Saturn and its perturbations on the Trojan swarms.


\subsubsection{\textsc{10embryos}}

In order to complete the study of the embryo case, we decided to do a small statistical study about the probability of losing massive Trojans from the swarms. Since simulations with almost hundred massive bodies are computational expensive, we decided to run 10 simulations with Jupiter and Saturn's growth tracks and substituting only 10 of their Trojans as Ceres-like bodies in a random way, that is not choosing in which swarm they end up. 
\begin{table}
\caption{Planetary embryos left in each swarm in the \textsc{10embryo} simulations}             
\label{table:10embryos}      
\centering          
\begin{tabular}{c |c |c |c |c }     
\hline\hline       
Run&Starting L$_4$&Starting L$_5$&L$_4$ Trojan&L$_5$ Trojan\\
&Trojans&Trojans&left&left\\
\hline                    
 Run1&5&5&1&0\\
 Run2&4&6&1&1\\
 Run3&8&2&2&0\\
 Run4&5&5&1&0\\
 Run5&6&4&1&0\\
 Run6&6&4&1&2\\
 Run7&8&2&1&0\\
 Run8&5&5&2&1\\
 Run9&7&3&1&1\\
 Run10&5&5&1&0\\
\hline                 
\end{tabular}
\end{table}
In Table \ref{table:10embryos} we display the name of the run in the first column, the starting L$_4$ massive Trojans in column two, the starting L$_5$ massive Trojans in column three and the final number of embryos after 4.5 Gyr in L$_4$ and L$_5$ in columns four and five, respectively. In all the ten runs we ended up with at least one massive embryo left in one of the swarms. Even if in 6 runs we have no massive embryo surviving in the L$_5$ swarm, it seems that it is very hard to deplete both swarms in the same run.

We conclude that the hypothesis of embryos embedded in the Trojan swarms presents two main problems: (a) it is hard to get rid of the last embryos in the swarms and (b) the presence of embryos can heavily affect the original asymmetry ratio in decreasing it, increasing it and also reversing it.


\subsection{\rm{\textsc{Jprestirred}} and \rm{\textsc{JSprestirred}}}

In the last scenario, we tested the resulting Trojan orbital proprieties if they are captured from an already pre-stirred disc. 
The simulations are exactly the same as J0 and JS0, we just modified the small massless body inclinations and eccentricities in the disc. 

\subsubsection{\textsc{Jprestirred\_ab} and \textsc{JSprestirred\_ab}}

In the \textsc{Jprestirred\_ab} and \textsc{JSprestirred\_ab} simulations, we kept track of the number of Trojans and their asymmetry ratio during $t=4.5$ Gyr of evolution. 

\begin{table}
\caption{Evolution of the number of Trojans, their mass and asymmetry ratio for simulations \textsc{Jprestirred\_ab}, \textsc{JSprestirred\_ab}, \textsc{Jprestirred\_ckbo} and \textsc{JSprestirred\_ckbo}.}             
\label{table:Jab}      
\centering          
\begin{tabular}{c |c |c |c |c}     
\hline\hline       
Time&Trojans&Mass&Trojan&Asymmetry\\
(Myr)&left&left (M$_\oplus$)&depletion (\%)&ratio\\

\multicolumn{5}{c}{\textsc{Jprestirred\_ab}}\\
\hline                    
5&2855&0.14&0.0&1.15\\
1000&2709&0.14&5.1&1.10\\
4500&2684&0.13&6.0&1.11\\
  
\multicolumn{5}{c}{\textsc{JSprestirred\_ab}}\\
\hline
5&760&0.038&0.0&1.28\\
1000&269&0.013&64.6&1.28\\
4500&173&0.009&77.2&1.14\\

\multicolumn{5}{c}{\textsc{Jprestirred\_ckbo}}\\
\hline                    
5&2643&0.13&0.0&1.15\\
1000&2556&0.13&3.3&1.11\\
4500&2544&0.13&3.7&1.11\\
  
\multicolumn{5}{c}{\textsc{JSprestirred\_ckbo}}\\
\hline
5&666&0.033&0.0&1.38\\
1000&223&0.011&66.5&1.25\\
4500&165&0.008&75.2&1.06\\

\hline                 
\end{tabular}
\end{table}


\begin{figure}
\begin{center}
\includegraphics[width=\hsize]{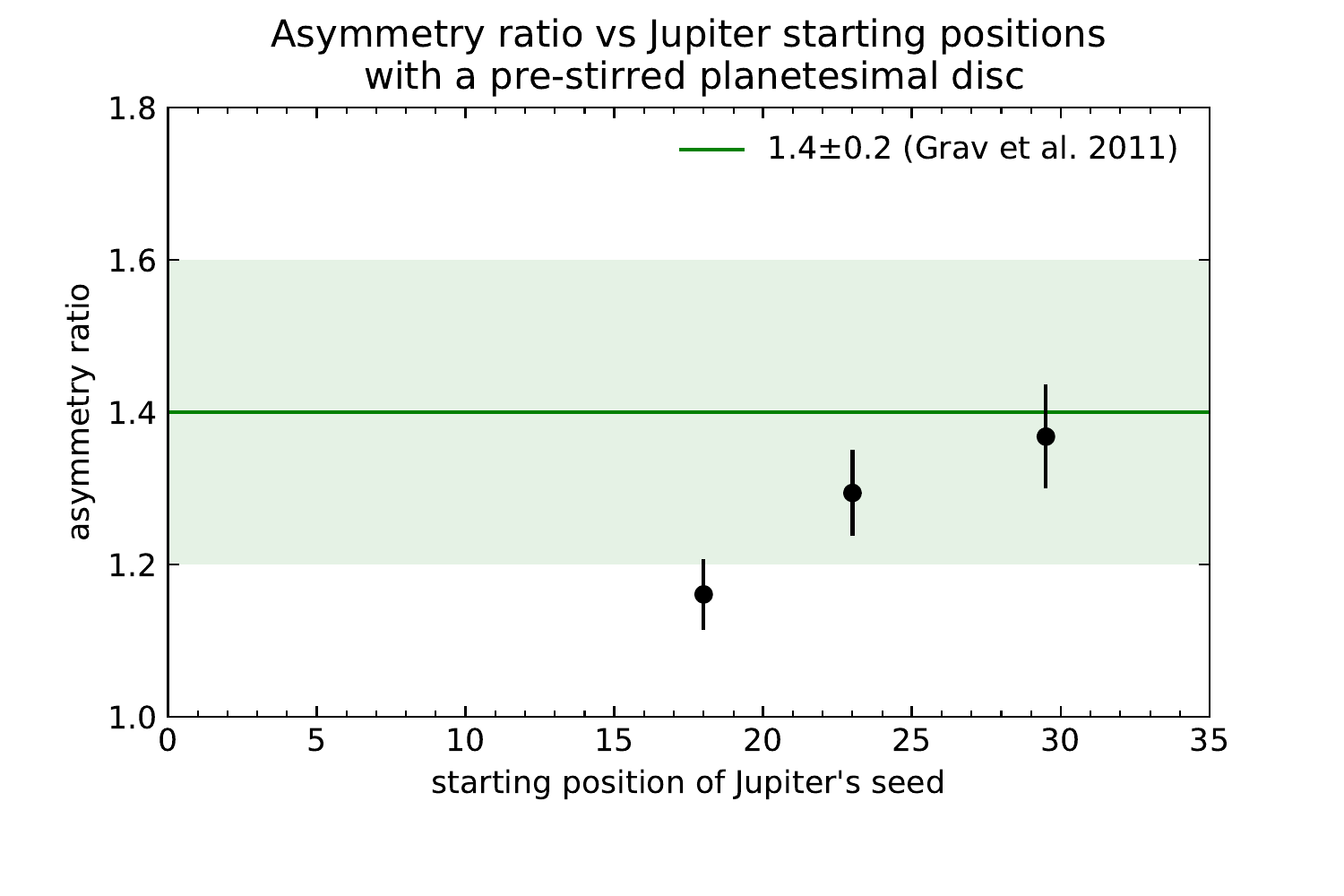}
\caption[]{The asymmetry ratio of the Jupiter Trojans obtained starting Jupiter's seed in different positions: 18 au (nominal case), 23 au and 29.5 au. This is the case with simulations with only Jupiter involved and an initial planetesimal disc stirred to values resembling the asteroid belt inclination and eccentricity distributions. The current asymmetry ratio of the Jupiter Trojans is highlighted in green.}
\label{figure:faster}
\end{center}
\end{figure}

In the case where only Jupiter is present in the system (\textsc{Jprestirred\_ab}), we can see from Table \ref{table:Jab} that Jupiter captures a number of Trojans of the same order of the one obtained in J0. Also, the depletion of the mass is not effective over $t=4.5$ Gyr as in J0. The asymmetry, instead, is smaller than in \textsc{J0} and remains similar to the initial one. 
The low ratio is due to less efficiency in the mechanism generating the asymmetry in the pre-stirred case with only Jupiter involved. In fact, when we experimented with letting Jupiter start also at 23 and 29.5 au, we obtained asymmetries of about 1.3 and 1.4 as shown in Figure \ref{figure:faster}, respectively. We computed the arithmetic mean of the values found in the 10 simulations for each case and the uncertainty is represented by the unbiased standard deviation. The current asymmetry ratio is also highlighted in green. For the asymmetry ratio error, a propagation of the uncertainty is applied. 

If we analyse the case where also Saturn is added to the system (\textsc{JSprestirred\_ab}), we notice that the number of particles trapped as Trojans is much less compared to \textsc{Jprestirred\_ab} simulations. Probably, again perturbations exerted by Saturn on Jupiter Trojans are very effective on already pre-stirred disc particles. The final mass of the Trojans is of the order of $10^{-2}$ M$_\oplus$, still significantly larger that the current Trojan's mass that is roughly $10^{-5}$ M$_\oplus$ \citep{vinogradova15}, but we need to account for an extra depletion due to the late instability of the giant planets, as already discussed in section \ref{sec:methods}. The asymmetry is also smaller compared to the flat disc case, but remains of the same order. At $t=4.5$ Gyr only few Trojans are left in the swarms and fluctuations in the depletion history can affect heavily the asymmetry, so the last number is not particularly meaningful. 

\begin{figure}
\begin{center}
\includegraphics[width=\hsize]{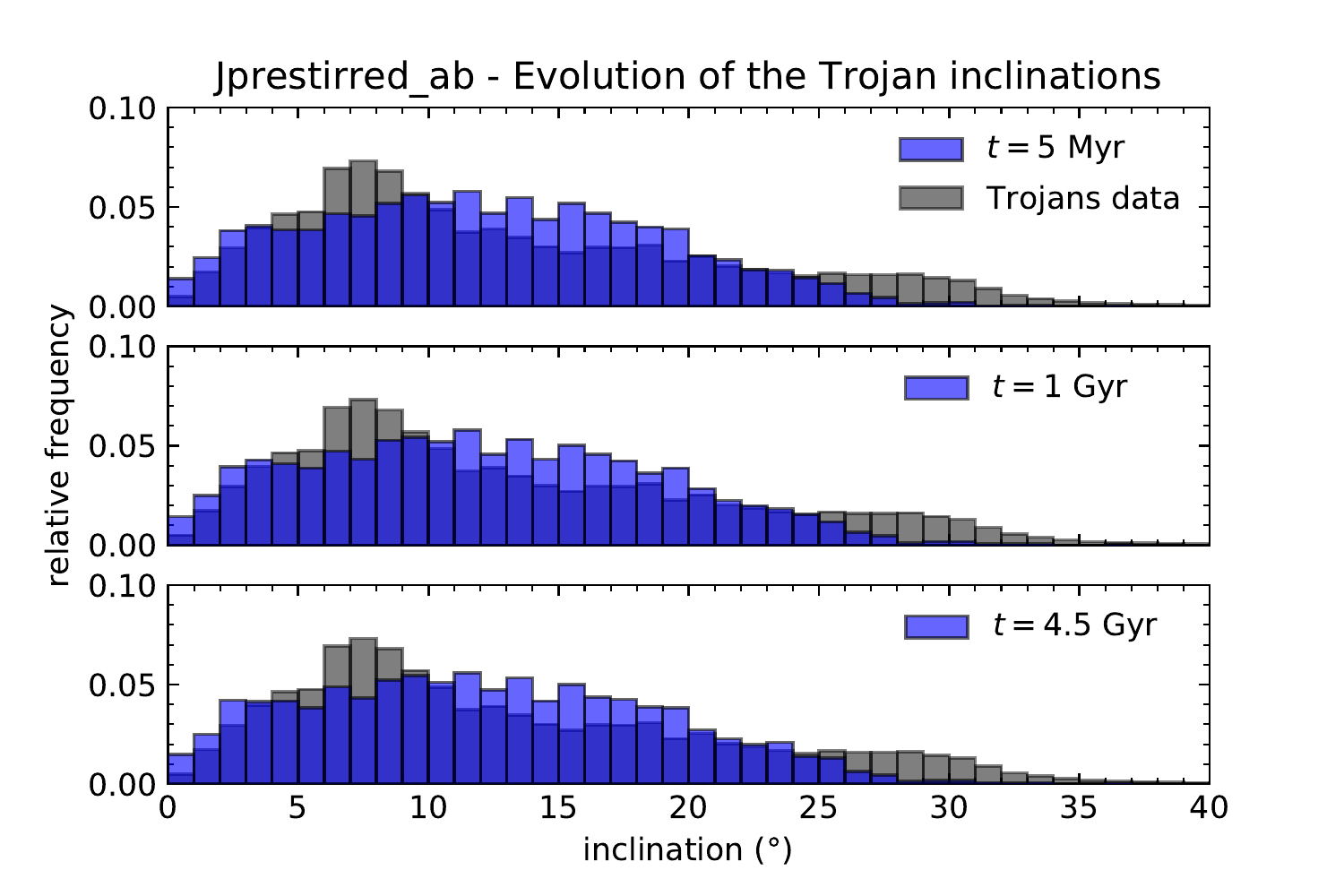}
\includegraphics[width=\hsize]{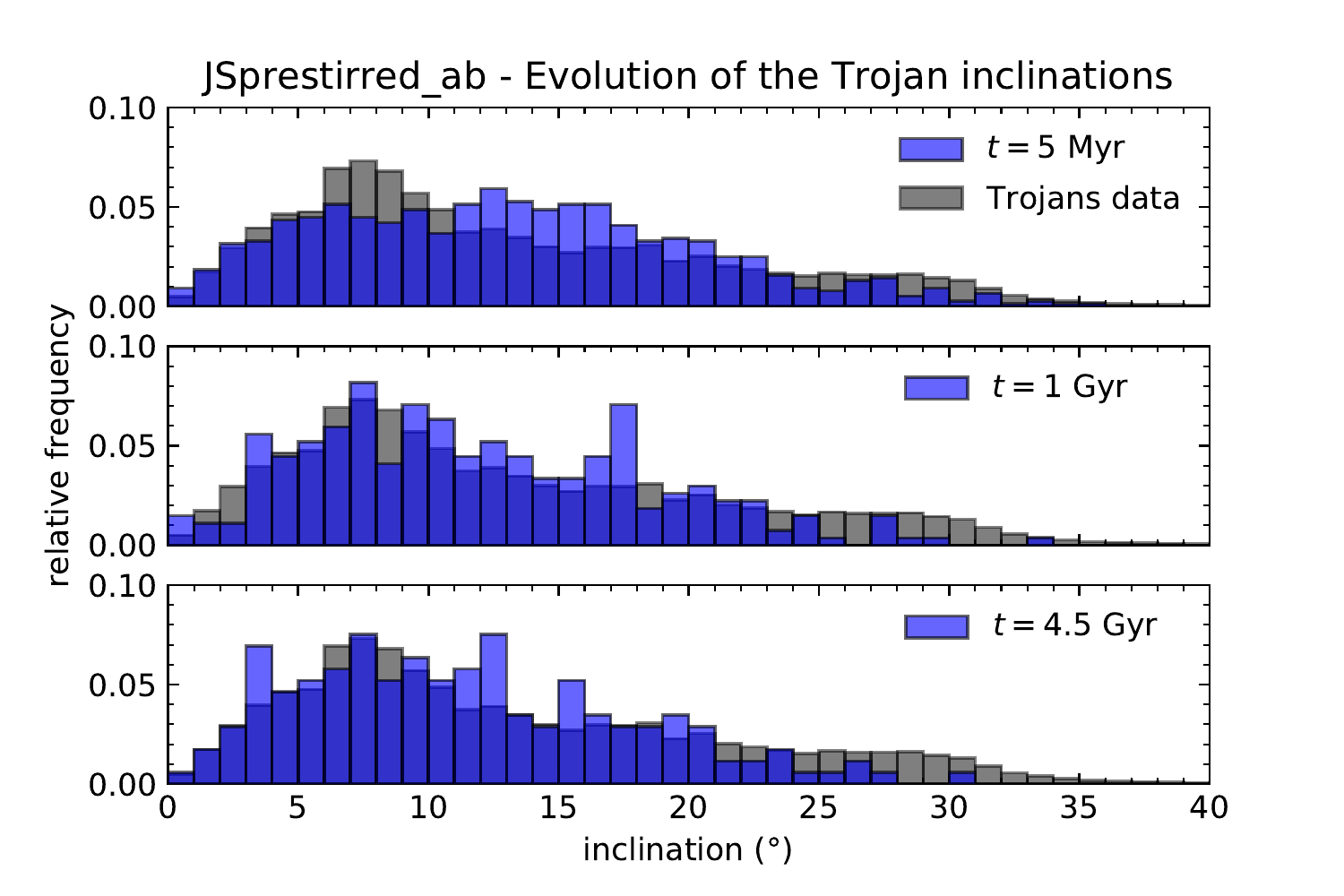}
\caption[]{Top figure: inclination distribution of the Jupiter Trojans at $t=5$ Myr (top histogram), $t=1$ Gyr (middle histogram) and $t=4.5$ Gyr (bottom histogram) from simulation \textsc{Jprestirred\_ab}. Bottom figure: inclination distribution of the Jupiter Trojans at $t=5$ Myr (top histogram), $t=1$ Gyr (middle histogram) and $t=4.5$ Gyr (bottom histogram) from simulation \textsc{JSprestirred\_ab}.}
\label{fig:Jabi}
\end{center}
\end{figure}

The inclination distribution of the Jupiter Trojans in simulation \textsc{Jprestirred\_ab} are shown in Figure \ref{fig:Jabi}, top panel. In blue is displayed the Trojan inclinations from the simulations at $t=5$ Myr (top histogram), $t=1$ Gyr (middle histogram) and $t=4.5$ Gyr (bottom histogram). Aerodynamic gas drag is not effective in the outer Solar System while the Trojans are trapped. Particles spend 2 Myr in the disc before Jupiter starts to grow, but in this time-span, eccentricities and inclinations are not damped enough to get a flat disc. The shape of the inclination distribution, instead, is preserved in the Trojan population one. If we analyse the results at $t=5$ Myr, we notice that the quasi-invariance of the inclinations of the Jupiter Trojans under mass growth and inner migration of the gas giant \citep{fleming00} holds also for the higher inclinations of the pre-stirred disc.  
The same happens in the case in which we add Saturn to the system as shown in Figure \ref{fig:Jabi}, bottom panel: disc particles preserve the high inclination distribution in the captured Jupiter Trojan population, but the presence of Saturn also shapes it in a way that is more similar to the current inclination distribution of the Trojans. 

\subsubsection{\textsc{Jprestirred\_ckbo} and \textsc{JSprestirred\_ckbo}}

\begin{figure}
\begin{center}
\includegraphics[width=\hsize]{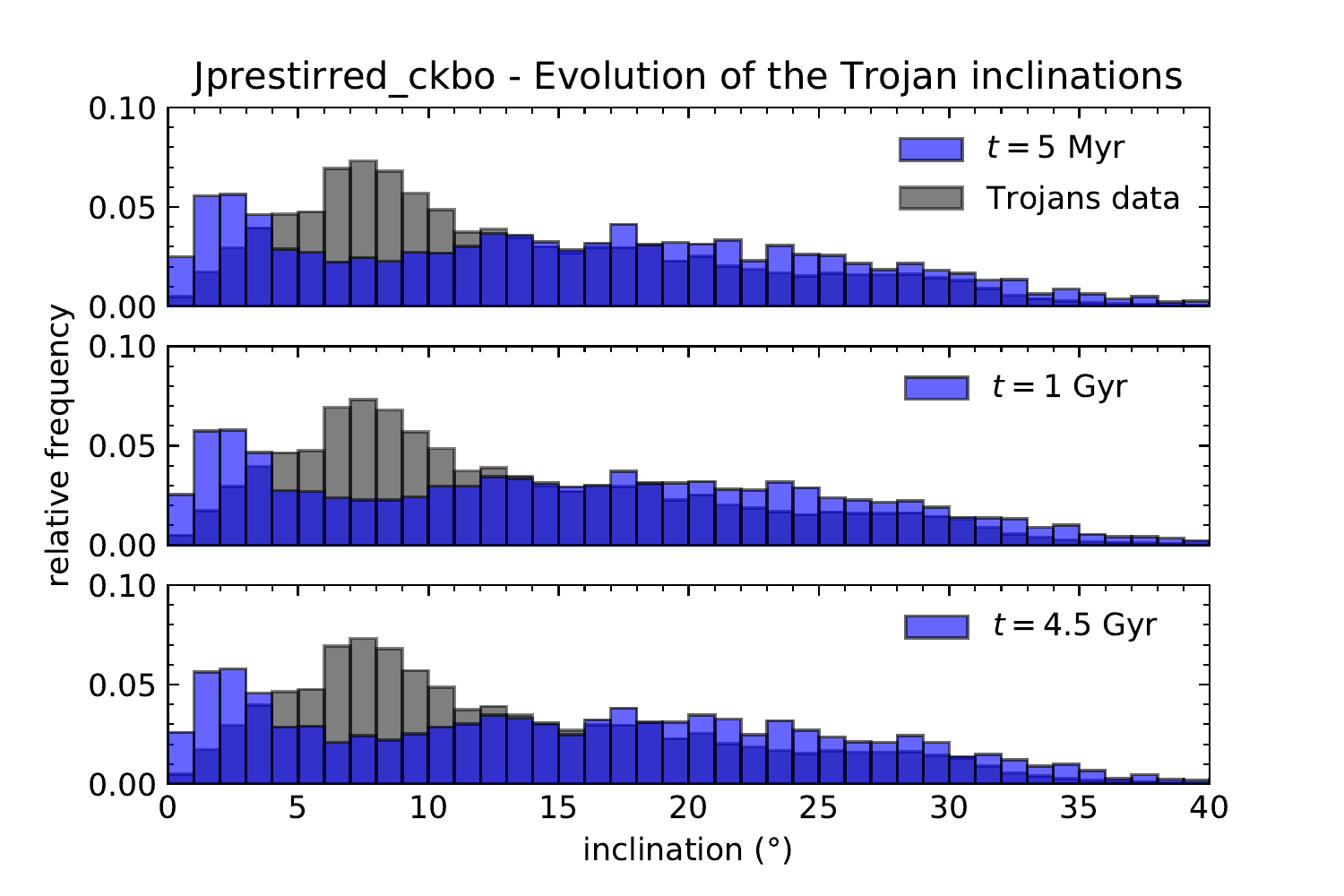}
\includegraphics[width=\hsize]{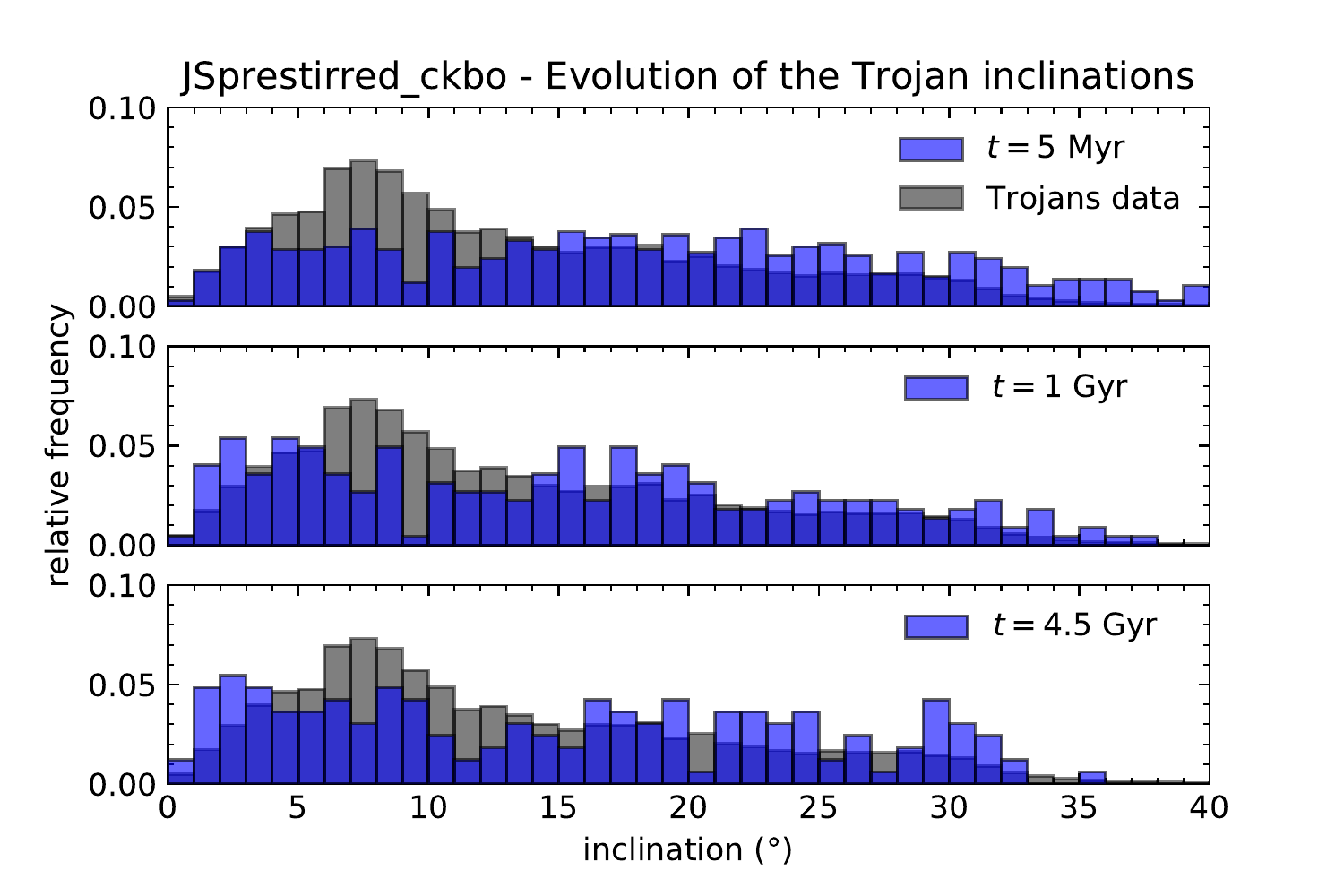}
\caption[]{Top figure: inclination distribution of the Jupiter Trojans at $t=5$ Myr (top histogram), $t=1$ Gyr (middle histogram) and $t=4.5$ Gyr (bottom histogram) from simulation \textsc{Jprestirred\_ckbo}. Bottom figure: inclination distribution of the Jupiter Trojans at $t=5$ Myr (top histogram), $t=1$ Gyr (middle histogram) and $t=4.5$ Gyr (bottom histogram) from simulation \textsc{JSprestirred\_ckbo}.}
\label{fig:Jcci}
\end{center}
\end{figure}

In the \textsc{Jprestirred\_ckbo} and \textsc{JSprestirred\_ckbo} simulations, we kept track of the number of trapped Trojans and their asymmetry ratio during $t=4.5$ Gyr of evolution, as in the previous cases. We reported the data in Table \ref{table:Jab}. 
As in \textsc{Jprestirred\_ab} the Trojans captured in simulation \textsc{Jprestirred\_ckbo} are of the same order of the flat disc one with only Jupiter (\textsc{J0}); the depletion in mass is also very small and the asymmetry ratio is and follows the evolution of simulation \textsc{Jprestirred\_ab}.
Same analogies in between simulation \textsc{JSprestirred\_ckbo} and \textsc{JSprestirred\_ab}: less particles captured as Trojans, significant depletion of the swarms and an asymmetry ratio that remains of the same order.
The inclination distribution of \textsc{Jprestirred\_ckbo} preserves the initial shape as happened for \textsc{Jprestirred\_ab}, as shown in figure \ref{fig:Jcci}, top panel. In the case in which also Saturn is added (\textsc{JSprestirred\_ckbo}), instead, Trojans remain very inclined, but the initial shape is not recognisable (Figure \ref{fig:Jcci}, bottom panel), as happens for \textsc{JSprestirred\_ab}, too.
For the final total mass of the Trojans in these simulations, we obtained similar results as in the \textsc{Jprestirred\_ab} and \textsc{JSprestirred\_ab} cases.


\section{Discussions and conclusions}\label{sec:conclusions}

In this paper, we tested different possibilities in order to explain the high inclination distribution observed in the Jupiter Trojans: (1) secular evolution of an initially flat Trojan population, (2) embryos embedded in the Trojan swarms and (3) pre-stirred planetesimals trapped as Trojans. All of our simulations are based on the core accretion model boosted by pebble accretion that allow the cores of the giant planets to grow fast enough to accrete gas from the protoplanetary disc. While the protoplanets are growing, they also experience inward migration because of the interactions with the surrounding gas. The resulting scenario is a large scale migration of a growing Jupiter until the gas of the protoplanetary disc is still available, that is in our case 3 Myr. Based on our results, our main conclusions are:
\begin{enumerate}[(a)]

\item If Jupiter Trojans are captured from a disc of particles with zero inclinations, their secular evolution does not result in any significant increase of their inclinations. The inclination distribution will remain very flat during the 4.5 Gyr of evolution of the Solar System. The system is also very stable in the case in which only Jupiter is in the system, that is there is no depletion of the Trojans' mass. In the case in which also Saturn is added to the system, perturbations between the planets will lead to a less efficient capture (15\% less), a slight increase of the inclinations of the Trojans and to a depletion of the swarms of the order of about 40\%. It is, however, very unlikely that the planetesimal disc in which planetary embryos are forming could have remained unaffected by the presence of massive bodies and hence remain very flat when Jupiter was growing.

\item If massive planetesimals are embedded in the Trojan swarms, then the inclination distribution of the Trojans can evolve to agree with the current one, especially if the embryos are at least of similar mass to Pluto. Moreover, embryos are very effective in depleting the swarms. It is important to notice that an uneven distribution of the embryos in between the two swarms can affect the original asymmetry and can even reverse it. This means that in the massive embryos model, the current observed asymmetry in the Jupiter Trojans might not reflect the initial one generated in the early inward migration and growth of the gas giant. The main problem is represented by the necessity of getting rid of the embryos during the 4.5 Gyr of evolution of the Solar System since in the current Trojan population there is no very massive asteroid. In none of our set of simulations did we successfully lose all the embryos from both the swarms. When one of the swarms is left with just one embryo, embryo--embryo scattering is not possible anymore and the massive object's probability of being lost would be as that of a massless Trojan: it would just depend on its eccentricity and inclination \citep{levison97}.  

\item When we considered a pre-stirred planetesimal disc and added Saturn to the simulations, our captured Trojan population is less massive than the flat case by an order of magnitude. The subsequent depletion also accounts for another order of magnitude in the mass loss. Finally, the late instability would at least account for another substantial depletion, that is at least 80\% as found by \citet{pirani19}. The resulting Trojans preserve the initial high inclination distribution and it is very similar to the current one. The asymmetry in the cases where only Jupiter is migrating is lower compared to the other cases and it is to be attributed to less efficiency in generating the asymmetry. Starting Jupiter slightly further away from the Sun generates an asymmetry consistent with the observed one.

\end{enumerate}

We want to stress that the estimate of the mass of each particle, and hence the mass of the Trojan population, is based on the minimum mass solar nebula, a model that does not necessarily represent the primordial planetesimal populations, since the planets migrate in the disc during their formation. Hence, any computation of the Trojans mass' absolute value is to be taken with a grain of salt. Moreover, the late instability of the giant planets has not been simulated in this paper, since the mechanism, time and time-scales of this event are still uncertain. A good fraction of the Trojan's mass is probably lost in this event as shown by \citet{pirani19} who estimated that only roughly 20\% of the Jupiter Trojans survive if the giant planet jumps suddenly from 5.4 to 5.2 au. 
We conclude that a pre-stirred planetesimal disc is the most likely scenario for the Trojans' capture, since this can explain simultaneously the high inclinations, the low total mass and the asymmetry ratio of the Jupiter Trojans.


\begin{acknowledgements}

We would like to thank the anonymous referee for the helpful comments. SP, AJ and AM are supported by the project grant `IMPACT' from the Knut and Alice Wallenberg Foundation (grant 2014.0017). AJ was further supported by the Knut and Alice Wallenberg Foundation grants 2012.0150 and 2014.0048, the Swedish Research Council (grant 2018-04867) and the European Research Council (ERC Consolidator Grant 724687- PLANETESYS). The computations are performed on resources provided by the Swedish Infrastructure for Computing (SNIC) at the LUNARC-Centre in Lund and partially funded by the Royal Physiographic Society of Lund through a grant.  
    
\end{acknowledgements}

%
   \bibliographystyle{aa} 
   \bibliography{pirani} 
%


\end{document}